\documentclass{aa}

\usepackage[varg]{txfonts}
\usepackage{graphicx}
\usepackage{microtype}
\usepackage[english]{babel}

\usepackage{natbib}
\bibpunct{(}{)}{;}{a}{}{,} 

\newcommand{\mps}{m\,s$^{-1}$}
\newcommand{\dqinv}{\delta q_{\alpha}^{\mathrm{inv}}}
\newcommand{\akern}{\mathcal{K}^{\alpha}_{\beta}}
\newcommand{\akerninv}{\mathcal{K}^{\alpha}_{\alpha} \left(\vec{r'}, z; z_0\right)}
\newcommand{\lab}{\tens{\Lambda}^{ab}}
\newcommand{\w}{w^{\alpha}}
\newcommand{\dq}{\delta q^{\beta}}
\newcommand{\ssp}{sound-speed perturbations}
\newcommand{\dcs}{\delta c_s}
\newcommand{\kw}{\mbox{$k\text{--}\omega$}}

\newcommand{\dif}[1]{\mathrm{d}{#1}\,}
\newcommand{\ddif}[1]{\mathrm{d}^2\vec{{#1}}\,}
\newcommand\norm[1]{\left\lVert#1\right\rVert}
\newcommand{\FWHM}{\textnormal{FWHM}}

\newcommand{\ocfigure}[4][tbp]{
    \begin{figure}[#1]
        \resizebox{\hsize}{!}{\includegraphics{#2}}
        \caption{#3}
        \label{#4}
    \end{figure}
}

\newcommand{\tcfigure}[4][tbp]{
    \begin{figure*}[#1]
        \centering
        \includegraphics[width=17cm]{#2}
        \caption{#3}
        \label{#4}
    \end{figure*}
}

\newcommand{\wfigure}[4][tbp]{
    \begin{figure*}[#1]
        \sidecaption
        \includegraphics[width=12cm]{#2}
        \caption{#3}
        \label{#4}
    \end{figure*}
}

\begin{document}

\title{
Plasma flows and \ssp{}\\in the average supergranule
}

\author{
    David Korda\inst{1}
    \and
    Michal {\v S}vanda\inst{1,}\inst{2}
}

\offprints{David Korda, \\ \email{korda@sirrah.troja.mff.cuni.cz}}

\institute{
    Astronomical Institute of Charles University, Faculty of Mathematics and Physics, V~Hole\v{s}ovi\v{c}k\'ach 2, Praha 8, CZ-180~00, Czech Republic
    \and
    Astronomical Institute of Czech Academy of Sciences, Fri\v{c}ova 298, Ond\v{r}ejov, CZ-25165, Czech Republic
} 

\abstract
{Supergranules create a peak in the spatial spectrum of photospheric velocity features. Even though they have some properties of convection cells, their origin is still being debated in the literature. The time--distance helioseismology constitutes a method that is suitable for investigating the deep structure of supergranules. }
{Our aim is to construct the model of the flows in the average supergranular cell using fully consistent time--distance inverse methodology. }
{We used the Multi-Channel Subtractive Optimally Localised Averaging inversion method with regularisation of the cross-talk. We combined the difference and the mean travel-time averaging geometries. We applied this methodology to travel-time maps averaged over more than $10^4$ individual supergranular cells. These cells were detected automatically in travel-time maps computed for 64 quiet days around the disc centre. The ensemble averaging method allows us to significantly improve the signal-to-noise ratio and to obtain a clear picture of the flows in the average supergranule. }
{We found near-surface divergent horizontal flows which quickly and monotonously weakened with depth; they became particularly weak at the depth of about 7~Mm, where they even apparently switched sign. The amplitude of the `reversed' flow was comparable to the background flows. The inverted vertical flows and \ssp{} were spoiled by unknown systematic errors. To learn about the vertical component, we integrated the continuity equation from the surface. The derived estimates of the vertical flow depicted a sub-surface increase from about 5~\mps{} at the surface to about 35~\mps{} at the depth of about 3~Mm followed by a monotonous decrease to greater depths. The vertical flow remained positive (an upflow) and became indistinguishable from the background at the depth of about 15~Mm. We further detected a systematic flow in the longitudinal direction. The course of this systematic flow with depth agrees well with the model of the solar rotation in the sub-surface layers. }
{}

\keywords{
    Sun: helioseismology -- Sun: oscillations -- Sun: interior
}

\authorrunning{D. Korda and M. {\v S}vanda}

\maketitle


\section{Introduction}

The spatial spectrum of velocity features in the solar photosphere is continuous with two clearly visible distinct peaks. The broad peak located around the angular degree of about 1500 is related to granulation. The granules constitute the top-most convective mode, which corresponds to the thermal dissipation scale. The sharp peak at the angular degree of about 120 corresponds to so-called supergranules. 

Supergranules were discovered by \citet{hart_1954} and have been intensively studied since that time. A textbook explanation of the origin of the supergranules is connected with thermal instability. A recombination of \ion{He}{iii} to \ion{He}{ii} at around 10~Mm below the solar surface releases latent heat \citep{Gierasch_1985}. The latent heat triggers thermal instability, which can lead to motions at the supergranular scale (about 20--30~Mm). However, this only holds true for fluid in rest. The convection zone is naturally turbulent. The turbulent motion can suppress thermal instability. The state-of-the-art models of the solar convection \citep[e.g.][]{Rempel_2014} do not fully reproduce all properties of the supergranules. All the models predict upflows at the centre of a supergranule followed by divergent horizontal flows at the surface and downflows at the edge of the supergranule. A small increase in temperature at the centre of the supergranule is predicted as well. 

There are also other theoretical models considered in the explanation of the supergranular peak in the velocity spectrum; see, for example, a review by \cite{Hanasoge_2014}. Although the origin of the supergranules is still unclear, many papers were written about their properties, especially about plasma flows in a supergranule \citep{Svanda_2012, Duvall_2013, Duvall_2014, DeGrave_2015, Langfellner_2015, Bhattacharya_2017, Ferret_2019}, to name a few recent studies.

At the solar surface, the supergranules form places with strong horizontal divergent flows. The divergent \emph{horizontal} flows are clearly visible in Dopplergrams away from the disc centre. Around the centre of the solar disc, the \emph{vertical} supergranular flows can be measured from the Dopplergrams. The first successful measurements of the surface vertical flows were published by \citet{Duvall_2010}. They measured upflows of 10~\mps{} at the centre of the average supergranule and downflows of 4~\mps{} at the distance of about 14~Mm away from the centre of the average supergranule. The \ssp{} can be studied via intensity (or temperature) changes. The intensity contrast between the centre and the edge of the supergranule was successfully observed, for example, by \citet{Meunier_2007} or \citet{Goldbaum_2009}. \citet{Langfellner_2016} observed the intensity contrast of $\left(7.8 \pm 0.6\right) \times 10^{-4}$. The corresponding temperature changes were $1.1 \pm 0.1$~K.

The divergent flows highly affect travel times of waves. The `outflow--inflow' travel times have local minima at the centre of the divergent flows. Therefore, time--distance local helioseismology \citep{Duvall_1993} seems to be very useful in studying supergranules. The time--distance method interprets travel times of sound and surface gravity waves. These waves are generated via the vigorous convective motions on the solar surface and propagate through the solar interior. The travel time of the wave packet is sensitive to plasma and state parameters along the trajectory of the wave packet, such as temperature, density, plasma flows, and so on. This method was successfully used for inversions for various quantities, for example, plasma flows \citep{Duvall_2000, Svanda_2013}, \ssp{} \citep{Kosovichev_1997, Couvidat_2006}, meridional circulation \citep{Giles_1997, Giles_1998}, differential rotation \citep{Giles_1998}, or the far-side imaging of active regions \citep{Lindsey_2000}.

In the last decades, a large number of articles have been written about the structure of supergranules. Authors have mostly  focused  on plasma flows, using two main approaches. The first is based on constructing many \mbox{3D} forward models and chosing the model that minimises the given functional \citep[e.g. the model that best fits the observations, e.g.][]{Duvall_2013, DeGrave_2015}. The second uses inverse modelling: a set of \mbox{2D} inverse models are computed at several depths; see for example \cite{Svanda_2012}. In the presented article, we follow this latter approach.


\section{Methodology}

\subsection{Travel times}

The time--distance method studies sound and surface gravity waves via travel-time measurements of the waves between the two points on the surface, $\vec{r}_1$ and $\vec{r}_2$. The travel time is the time lag $\tau$ which maximises the temporal cross-correlation $C$ of the observed signal $\psi$ (e.g. Dopplergrams) at the two points:
\begin{align}
    C \left( \vec{r}_1, \vec{r}_2, t \right) &= \frac{1}{T} \int \limits_{-\infty}^{\infty} \dif{t'} \psi \left( \vec{r}_1, t' \right) \psi \left( \vec{r}_2, t' + t \right),\\
    \tau \left( \vec{r}_1, \vec{r}_2 \right) &= \arg \max \limits_t \left[C \left( \vec{r}_1, \vec{r}_2, t \right) \right],
\end{align}
where $T$ is the duration of the observation. In time--distance helioseismology, the perturbed travel time $\delta \tau$ is usually used. The perturbed travel time is defined as a difference between the measured travel time $\tau_{\mathrm{msm}}$ and the modelled one $\tau_{\mathrm{model}}$:
\begin{equation}
    \delta \tau \left( \vec{r}_1, \vec{r}_2 \right) \equiv \tau_{\mathrm{msm}} \left( \vec{r}_1, \vec{r}_2 \right) - \tau_{\mathrm{model}} \left( \vec{r}_1, \vec{r}_2 \right).
\end{equation}

To reduce the realisation-noise contribution to the observed travel times, four averaging geometries are introduced. Their formulation is based on the isotropic propagation of the wave packets. The averages are taken between the point $\vec{r} = \vec{r}_1$ and the surrounding annulus. This annulus has radius $\Delta = \norm{\vec{r}_1 - \vec{r}_2}$ and the centre in $\vec{r}$. Such averaging is usually referred to as point-to-annulus (PtA) averaging. We distinguish two omnidirectional averaging geometries, the \emph{outflow-inflow} (denoted o-i) which is the difference between the travel time from $\vec{r}$ to the rim of the annulus and the travel time in the opposite direction, and the \emph{mean} geometry, which takes an average of these two travel times propagating in the opposite way. In order to maintain sensitivity to the wave trajectory, the travel times measured on the annulus are weighted by the cosine or the sine of the polar angle. Subsequently, the difference between the travel times is computed in the same way as for the o-i geometry. These averaging geometries are called the \emph{east-west} for the cosine (denoted e-w) and the \emph{north-south} for the sine (denoted n-s). The geometries o-i, n-s, and e-w belong to the \emph{difference} travel-time geometries. 

\subsection{Forward modelling}

In the forward modelling,  the effect of the changes in the plasma state parameters on the observables, such as the travel times, are investigated. The sensitivity of the observables to changes of the background-model parameters is usually expressed by the sensitivity kernel $K$ in seismic applications. The sensitivity kernels quantify the perturbation of the travel times $\delta \tau$ due to the perturbations $\dq$ of the model parameters. The sensitivity kernels are computed using a given background model \citep[e.g. model S;][]{Model_S} under a chosen approximation \citep[e.g. Born approximation;][]{GB02} for a specific perturbation (e.g. \ssp{} -- \citealt{B04} -- or vector flows -- \citealt{BG07}).

Using the sensitivity kernel, the PtA travel time $\delta \tau$ can be modelled using the equation
\begin{equation}
    \delta \tau^a \left(\vec{r}\right) = \int \limits_{\sun} \ddif{r'} \dif{z} \sum \limits_{\beta = 1}^P K^a_{\beta}\left(\vec{r'} - \vec{r}, z\right) \dq\left(\vec{r'}, z\right) + n^a\left(\vec{r}\right),
    \label{eq:dtau}
\end{equation}
where the upper index $a$ stands for the individual independent travel-time measurements and aggregates the selection of the mode filter, distance, and averaging. The coordinates $\vec{r}$ and $\vec{r'}$ are the horizontal positions, $z$ the vertical position, $P$ the number of the perturbed quantities in the model indexed by $\beta$, and $n^a$ represents the realisation noise of the given measurement.


\subsection{\mbox{MC-SOLA} inverse modelling}

The aim of inverse methods is to deconvolve Eq.~(\ref{eq:dtau}). This is very difficult as the noise term $n^a$ is usually the dominant term. Therefore, the inverse methods attempt to construct an estimate of $\delta q^{\alpha}$ denoted by $\dqinv$ in the general form:
\begin{equation}
    \dqinv = \dqinv \left( \delta \tau^a, K^a_{\beta}, n^a\right).
\end{equation}
The Multi-Channel Optimally Localised Averaging (\mbox{MC-SOLA}) method assumes that $\dqinv$ can be written as a linear combination of the perturbed travel times
\begin{equation}
    \dqinv \left( \vec{r}_0; z_0\right) = \sum \limits_{i = 1}^{N} \sum \limits_{a = 1}^{M} \w_{a} \left(\vec{r}_i - \vec{r}_0;z_0\right)\delta \tau^a\left(\vec{r}_i\right),
    \label{eq:inv_by_weights}
\end{equation}
where $\vec{r}_i$ and $\vec{r}_0$ are the horizontal positions, $N$ the number of considered horizontal positions, $z_0$ the target depth, and $\w_{a}$ the unknown weight functions which must be determined. The weight functions are determined by minimising the functional $\chi^2$ in the form:
\begin{align}
    \chi^2 &= \int \limits_{\sun} \ddif{r'} \dif{z} \sum\limits_\beta \left[\akern \left(\vec{r'}, z; z_0\right) - \mathcal{T}^{\alpha}_{\beta}\left(\vec{r'}, z; z_0\right)\right]^2 + \nonumber\\
    &+ \mu \sum \limits_{i,\,j,\,a,\,b} \w_a \left(\vec{r}_i; z_0\right) \lab \left(\vec{r}_i - \vec{r}_j\right)\w_b \left(\vec{r}_j; z_0\right) + \nonumber\\
    &+ \nu \sum \limits_{\beta \neq \alpha} \int \limits_{\sun} \ddif{r'} \dif{z} \left[\akern \left(\vec{r'}, z; z_0\right)\right]^2 +\epsilon \sum \limits_{a,\,i} \left[\w_a \left(\vec{r}_i; z_0\right)\right]^2 + \nonumber \\
    &+ \sum \limits_{\beta} \lambda^{\beta} \left[\int \limits_{\sun} \ddif{r'} \dif{z} \akern \left(\vec{r'}, z; z_0\right) - \delta^{\alpha}_{\beta}\right].
    \label{eq:chiSOLA}
\end{align}
\tcfigure{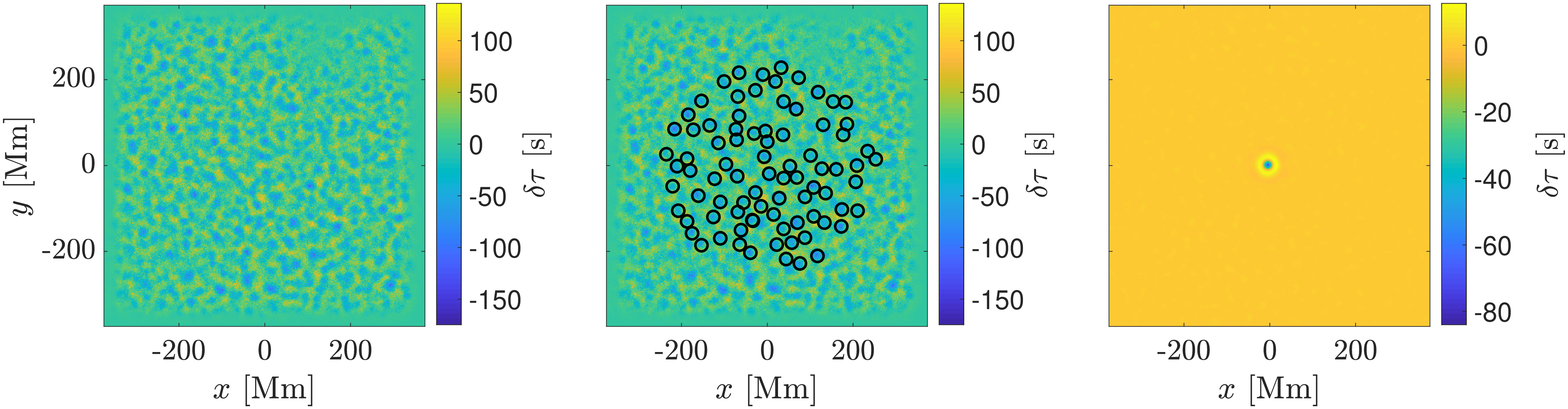}
{Example maps of $f$-mode travel time in the o-i geometry with $\Delta \approx 16$~Mm. Left: Measured travel time. Middle: Centres of the supergranules are labelled. Right: Corresponding travel time of the average supergranule. }
{fig:ave_sg}

The symbol $\mathcal{T}^{\alpha}_{\beta}$ indicates the user-selected target function localised around $z_0$, which serves as a user-given desire of the shape of the inversion averaging kernel (see below). Parameters $\mu$, $\nu,$ and $\epsilon$ are the trade-off parameters balancing the terms, $\lab$ represents the noise covariance matrices, $\lambda^{\beta}$ are the Lagrange multipliers bounding the constraint, and $\delta^{\alpha}_{\beta}$ is the Kronecker delta. For more details see \citet{Jackiewicz_2012} and \citet{Korda_2019a}. 

The function $\akern$ is termed the averaging kernel. It quantifies the level of smearing of the inverted quantity $\delta q^{\alpha}$, as well as the leakage of the other-than-wanted quantities $\delta q^{\beta \neq \alpha}$ into the inverted estimate $\dqinv$. The inverted estimate, the averaging kernels, and the true perturbations are connected as follows (compare to Eq.~(\ref{eq:inv_by_weights})):

\begin{align}
    \dqinv \left( \vec{r}_0; z_0\right) &= \sum\limits_{\beta} \int \limits_{\sun} \ddif{r'} \dif{z} \akern \left(\vec{r'} - \vec{r}_0, z; z_0\right) \dq \left(\vec{r'},z\right) + \nonumber \\
    &+ \sum \limits_{i, a} \w_{a} \left(\vec{r}_i - \vec{r}_0;z_0\right) n^a\left(\vec{r}_i\right).
    \label{eq:inv_by_akern}
\end{align}


\section{Data and data processing}
\subsection{Observed travel times}

For the travel-time measurements, we used SDO/HMI \citep{HMI1, HMI2} Dopplergrams of the photospheric iron line \ion{Fe}{i} 617.3~nm with the cadence of 45~s. We processed altogether 64 quiet days between 31 July 2010 and 23 February 2011. On each of the considered days, we tracked consecutive 12-hour datacubes with the Carrington rotation and selected the region of the disc centre 512$\times$512 pixels in the Postel's projection with a pixel size of 1.46~Mm and a cadence of 45~s. 

The Dopplergrams contain information about resonant standing waves trapped inside the Sun. Individual wave modes can be separated from each other in a \kw{} diagram by applying of a filter function. In this study, we used two types of filters, ridge filters, which separate resonant modes with the specific radial orders, and phase-speed filters, which select waves with similar trajectories. For the ridge filters, we selected 16 different radii $\Delta$ of the annuli; equidistantly from about 7.3~Mm to about 29~Mm. The standard phase-speed filters are defined by \citet{JSOC_pipeline}, for example. The travel times were measured in the linearised approximation presented by \citet{GB04}.


\subsection{Average supergranule}

Usually, the realisation noise is a significant component of the measured travel times due to the stochastic origin of the waves. The regularisation of the realisation noise is an important part of our inverse models. The noise regularisation always results in greater smearing of the inverted quantity. Moreover, the maximum depth sensitivity is always closer to the surface than the targeted depth; the averaging kernel deviates significantly from the target function. As a consequence of the greater smearing, the details are lost, especially the information about local amplitudes. \citet{DeGrave_2015} pointed out that the amplitude of the horizontal flows might be about 50\% lower when the smearing level is too great. \citet{Svanda_2012} constructed a simple model of supergranular flows. He demonstrated that the large-amplitude local flows are unobtainable with the time--distance method because of the smearing.

One way to circumvent this problem is to use the ensemble averaging method. This method uses a specific averaging over the different representatives of an abundant set of individual features of the same kind. This leads to noise suppression; relatively little regularisation of the noise terms is therefore needed in the inverse modelling and this in general results in less smearing and a better-quality averaging kernel.

The noise suppression is the main advantage of the ensemble averaging approach. Supergranules can be found on the whole surface. It is believed that the supergranules are a manifestation of stochastic convection and hence each quiet-Sun supergranule results from a random realisation of the same process. Therefore, the realisation noise is suppressed by a factor of $\sqrt{N}$ if $N$ independent supergranule realisations are averaged. 

The travel times of the average supergranule were computed in the following way \citep[inspired by][]{Svanda_2012}:

\begin{enumerate}
    \item We remove the known systematic trends for all the measured travel times (e.g. foreshortening in the case of the mean travel times); see Appendix~\ref{app:trend}.
    \item To detect the supergranular cells we use the $f$-mode o-i travel time with $\Delta \approx 16$~Mm.
    \item From these travel-time maps we remove small-scale structures by smoothing the travel time with a 3~Mm-\FWHM{} Gaussian filter.
    \item Then we label all points of the smoothed travel time whose values are smaller than the mean value of the smoothed travel time by more than two standard deviations. This indicates the regions where the supergranular centres may be expected. 
    \item Around each labelled point we segment out a squared neighbourhood (30~Mm on a side) in which we further keep only the point with the minimal travel time. By doing so for each compact region of the strongly negative travel time, we keep only its minimum. 
    \item We remove all the left-over points close to the edge of the field of view (less than 10\% away), and also, from the points closer than 23~Mm to each other, we keep only the point with a smaller travel time. In this step, we thus remove the unreliable points and also possible duplicates. The final set indicates the position of the clear supergranular cells. 
    
    \item Finally, for each geometry index $a$, we co-align all travel-time maps on the positions of supergranules and average them. 
\end{enumerate}

\noindent Figure~\ref{fig:ave_sg} shows an example of this process. In the left panel, there is the $f$-mode travel time in the o-i geometry with \mbox{$\Delta \approx 16$~Mm}. In the middle panel, the final centres of the supergranules are labelled after the shortlisting described above. In the right panel, the centres are co-aligned and averaged which results in the travel time of the average supergranule. In the analysed set of Dopplergrams, we found more than $10^4$ independent supergranule realisations and thus suppressed the realisation noise by a factor of about 100.


\section{Results}

To study the structure of the average supergranule, we computed two different sets of inverse models. The first set consisted of shallow inverse models with a rather large depth resolution, while the second set contained inverse models which scanned a large range of depths with a scarcer depth sampling.

\subsection{General properties of inversions}

The amplitude of the horizontal flows $\vec{v}_h = (v_x, v_y)$ is usually larger than the amplitudes of the vertical flows ($v_z$) and the \ssp{} ($\dcs$). Therefore, the inversions for the horizontal flows are not biased by the cross-talk. Also, the noise contribution is usually much smaller than the amplitude. For this reason, it is possible to invert for the horizontal flows down to the depth of about 5~Mm whilst maintaining reasonable localisation. In the case of the low-noise regime, we were able to achieve reasonable localisation down to a depth of about 17~Mm.

The average supergranule is a very symmetric object. Because of the large symmetry and the large amplitude of the horizontal flows, it is very important to regularise the cross-talk in the inversions for the vertical flows. Otherwise, the cross-talk leads to too much degradation of the inverted results. Although the realisation noise was suppressed via the ensemble averaging, the noise and the cross-talk regularisation worsen the localisations of the inversions. Reasonable localisation can be achieved down to a depth of about 3~Mm in the low-noise regime.

As well as in the case of the vertical flows, the amplitude of the \ssp{} is small compared to the horizontal flows. Therefore, the cross-talk must be regularised as well. In the low-noise regime, the inversions are well-localised down to a depth of about 4~Mm.


\ocfigure{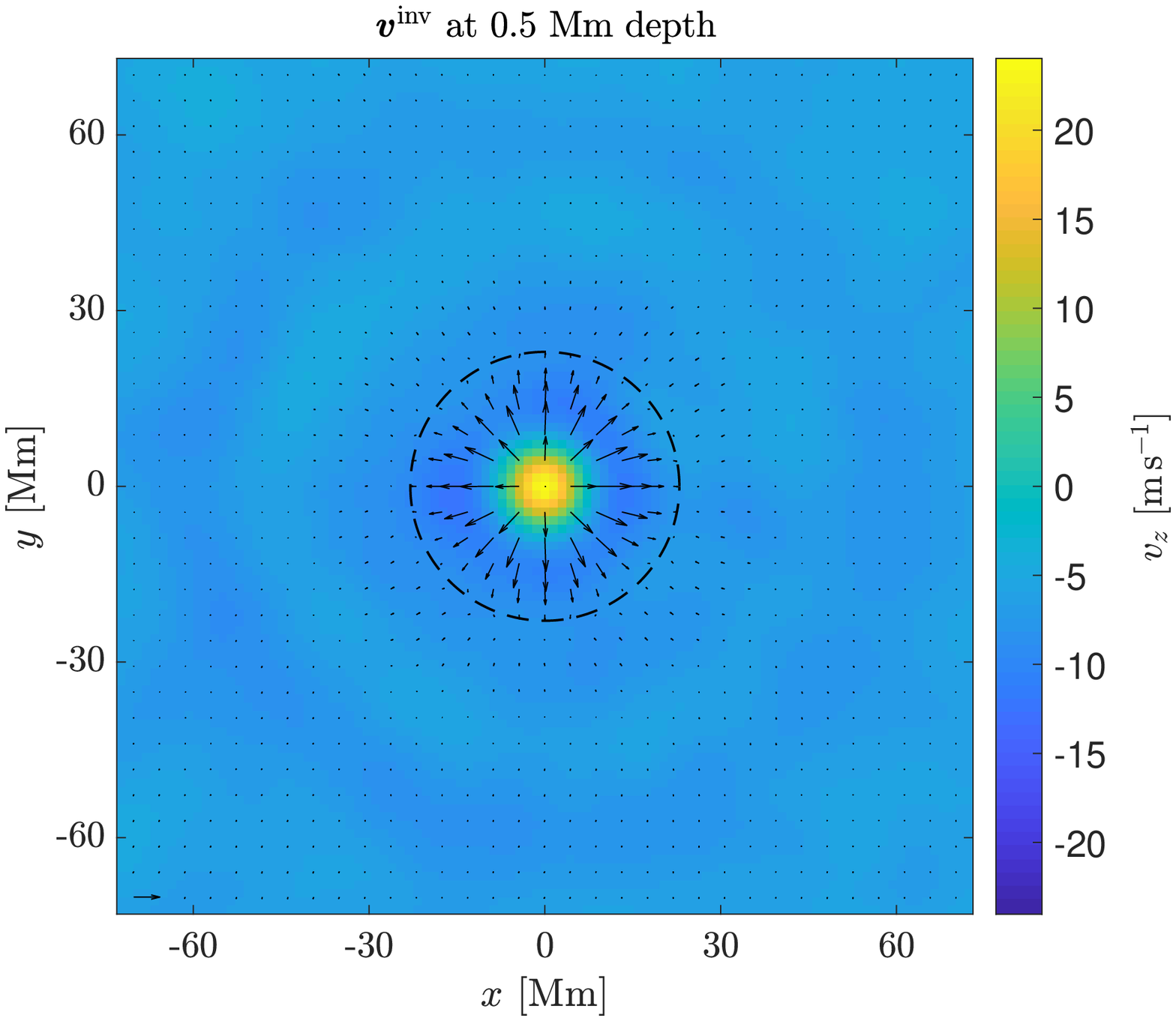}
{Inverted flows at 0.5~Mm depth. The colour map codes the vertical flows, where positive values correspond to upflows. The arrows correspond to the horizontal flows. The reference arrow in the bottom  left-hand corner corresponds to 250~\mps{}. The radius of the black dashed circle is approximately 23~Mm. }
{fig:flows_0.5}

\wfigure{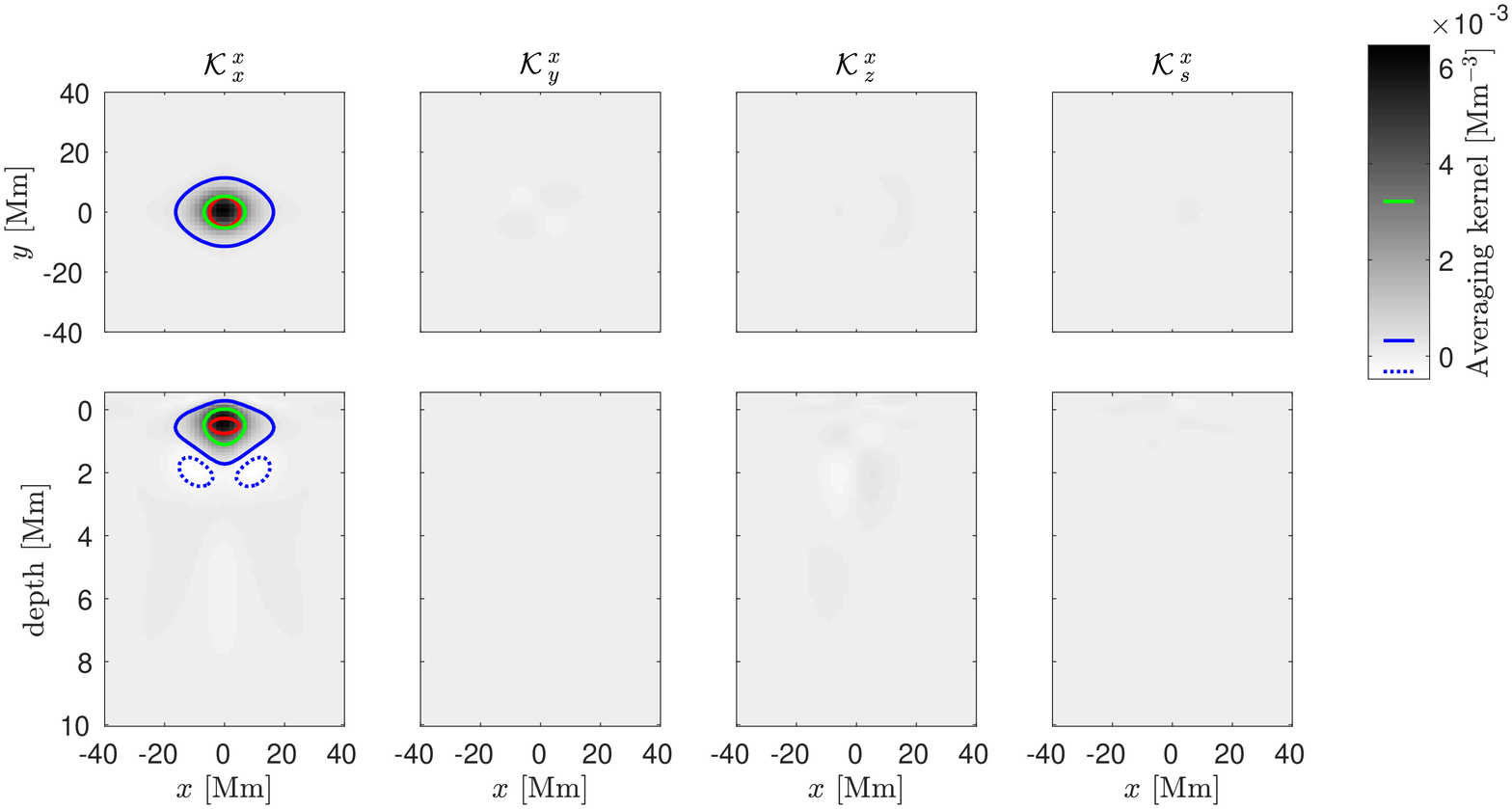}
{Averaging kernel for the inversion for the longitudinal flows at the target depth of 0.5~Mm. Top row: Horizontal cuts at the target depth. Bottom row: Vertical cuts at $y = 0$. Columns from left to right: Localisations of the longitudinal flows, latitudinal flows, vertical flows, and \ssp{}. The red curve corresponds to the half-maximum of the target function at the target depth. The green full, the blue full, and the blue dotted curves correspond to the half-maximum, $+5\%$, and $-5\%$ of the maximum of the averaging kernel at the target depth, respectively. }
{fig:vx_0.5_rakern}

\wfigure{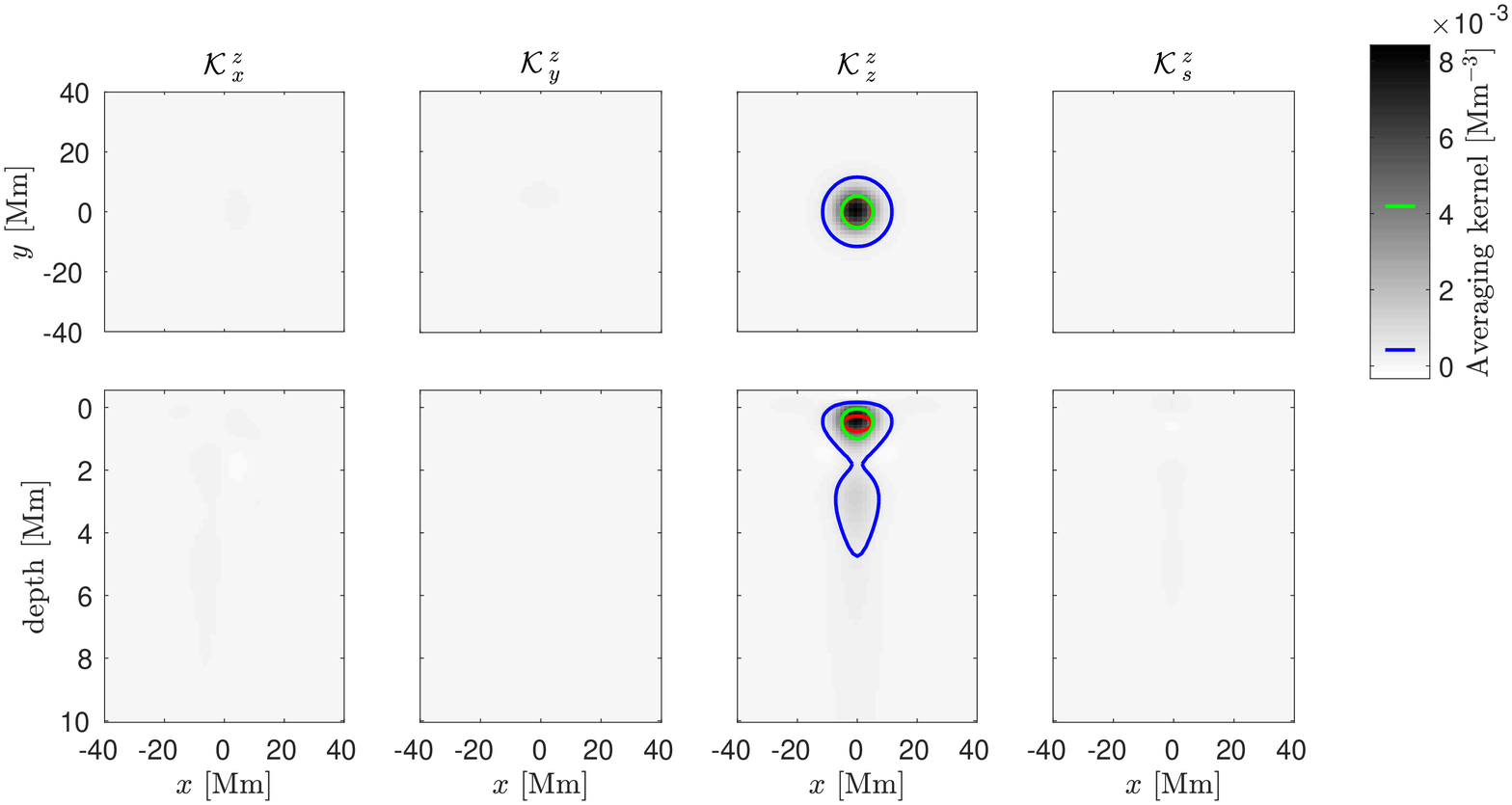}
{Averaging kernel for the inversion for the vertical flows at the target depth of 0.5~Mm. See the caption of Fig.~\ref{fig:vx_0.5_rakern} for details. }
{fig:vz_0.5_rakern}

\subsection{Surface inversions at the depth of 0.5~Mm}

\subsubsection{Vector flows}

The surface supergranular vector flows can be directly measured from the Dopplergrams. If the supergranule is far from the disc centre, we can observe the exhibition of the horizontal flows in the Dopplergrams, whereas the supergranules close to the disc centre contribute with the vertical-flow component to the Dopplergrams. The Dopplergrams indicate strong surface horizontal divergent flows and weak surface upflows at the supergranular centre compensated by weak surface downflows at the edge of the supergranule.

Our surface inversions for the vector flows are consistent with the observed Dopplergrams and the numerical simulations. The inverted supergranular vector flows are plotted in Fig.~\ref{fig:flows_0.5}. The centre of the average supergranule is at the coordinate $\left[0, 0\right]$. The colour map codes the vertical flows (positive values correspond to upflows and negative to downflows) and the arrows indicate the horizontal flows. The reference arrow in the bottom left-hand corner corresponds to 250~\mps{}. The black dashed circle indicates the edge of the average supergranule; its radius was computed as the distance at which the average latitudinal flows ($v_y$) change the sign. This radius was about 23~Mm which would result in a typical size of our average supergranule of about 46~Mm. This value is larger than what is usually given in the literature \citep[e.g.][]{Roudier_2014}. This apparent disagreement is a consequence of the detection algorithm, which prefers stronger divergent flows which are usually caused by larger supergranules. One can also see other circular upflow regions around the average supergranule. These correspond to supergranules which were next to the averaged supergranule. There is no preference in the position of the supergranules, and therefore this signal creates concentric circles of upflows smaller than the central upflows due to the averaging of the non-aligned neighbouring supergranules. 

The average supergranule is very symmetric around its centre. Some authors \citep[e.g.][]{2003Natur.421...43G,2003ApJ...596L.259S} pointed out that the supergranular pattern rotates faster than the plasma in supergranules. This was attributed to the wave-like component taking part in the formation of supergranules. This phenomenon has a negligible effect on our results. We do not track supergranules but invert for the speed of plasma flows. The slow prograde motion of supergranules with respect to the tracked background in principle could cause smearing of the flows in the longitudinal direction and perhaps a deformation of the average supergranular cell as derived from the flow pattern. We do not observe such asymmetry, and therefore effect of the supergranule `super-rotation' can  be safely neglected. The effects of super-rotation may appear in the supergranule power spectrum \citep{2018A&A...617A..97L}, which is not the goal of our study.

The localisations of the vector-flow inversions (only for $v_x$ and $v_z$ components, $v_y$ is similar to $v_x$) are visualised in Figs.~\ref{fig:vx_0.5_rakern} and \ref{fig:vz_0.5_rakern}. In the top rows, there are horizontal cuts at the target depth, while in the bottom rows, there are vertical cuts at $y = 0$. In both figures, the cross-talk components (columns two to four in Fig.~\ref{fig:vx_0.5_rakern} and one, two, and four in Fig.~\ref{fig:vz_0.5_rakern}) were negligible. The horizontal flows were localised around the target depth, while in the case of the vertical flows, the averaging kernel had a deeper lobe. The mean depth,
\begin{equation}
    \langle z \rangle \equiv \int \limits_{\odot} \ddif{r'} \dif{z} \akerninv z
,\end{equation}
of the vertical-flow inversion was about 1.8~Mm. The stronger upflows of about 25~\mps{} were a consequence of the deeper lobe. In Sect.~\ref{sec:deep_flows}, we show that the amplitude of the upflows at about 2-Mm depth is about 25~\mps{} which is in a great agreement with this inverse model.


\subsubsection{Sound-speed perturbations}

The small amplitudes made it impossible to invert for the \ssp{} with the separate inversion due to the presence of the cross-talk. We performed and validated the combined inversion for vector flows and \ssp{} \citep{Korda_2019a} with the cross-talk regularisation using the forward-modelled travel times which were based on a numerical simulation. 

However, here, in the case of the real-Sun inversions, the inverted results are not acceptable. In Fig.~\ref{fig:cs_0.5} we show inversions for the \ssp{} at 0.5-Mm depth for two different selections of the trade-off parameters. The inverted \ssp{} are completely different, not only in terms of their amplitude but also in their sign.

\wfigure{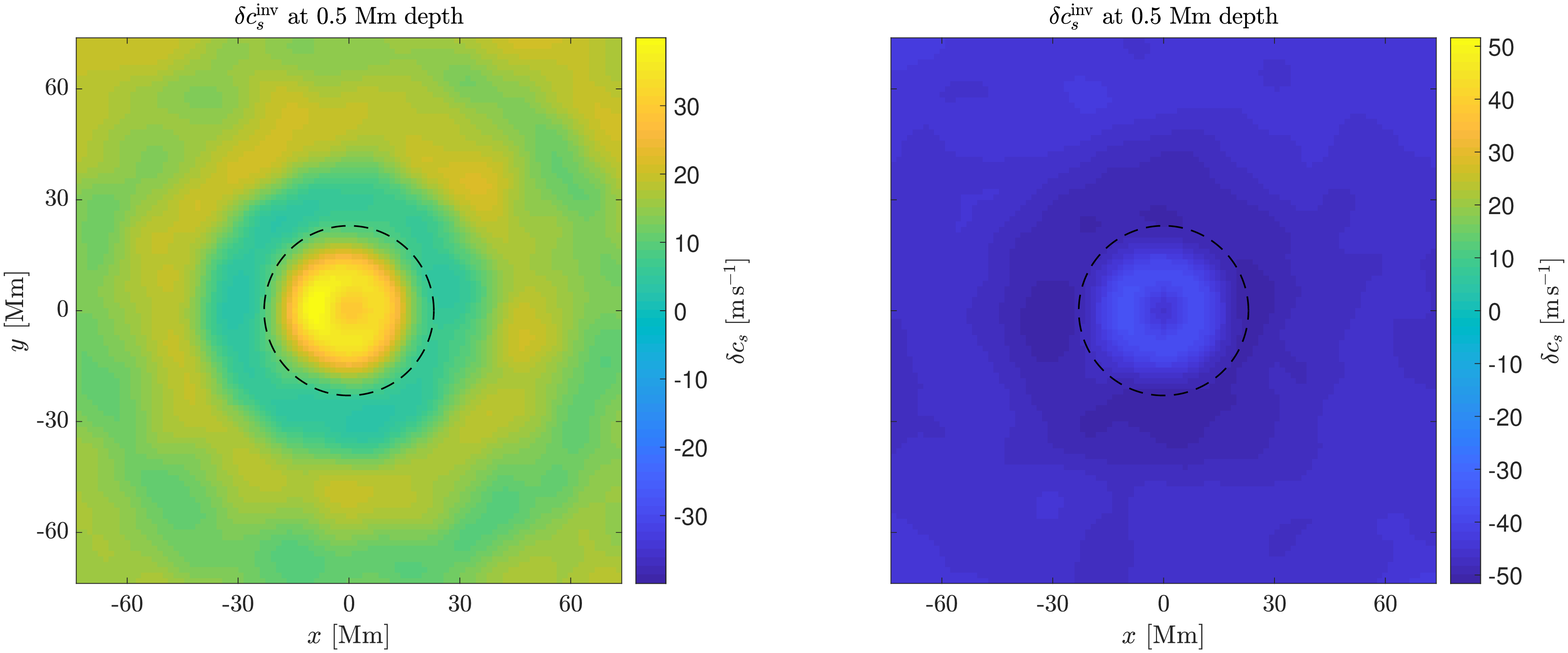}
{Inversions for the \ssp{} at 0.5~Mm depth. The sets of the trade-off parameters differ. The radii of the black dashed circles are approximately 23~Mm. }
{fig:cs_0.5}

These two inversions do not have to be inconsistent if the localisations of these inversions are different too. The corresponding averaging kernels are plotted in Figs.~\ref{fig:cs_0.5_rakern_ok} and~\ref{fig:cs_0.5_rakern_ko}. In both cases, the cross-talk components of the averaging kernels were negligible. Even though the inversion with the negative $\dcs$ at the supergranular centre has a deeper lobe, the localisations of both inversions are very similar. Therefore, the inverted results should be similar too, which unfortunately is not the case.  

\subsubsection{Search for the cause of inconsistency}

We found these issues in all inversions for the vertical flow and the \ssp{} for the target depths greater than 2~Mm. This issue is not new; analogous issues were discussed by \citet{Svanda_2015}, who inverted for the vector flows in the separate inversion setup. He found the issues in {all} the flow components. Therefore, the combined inversion used in this study returns consistent horizontal flows but does not improve the inversions for the vertical flows and \ssp{} around the average supergranule.

\wfigure{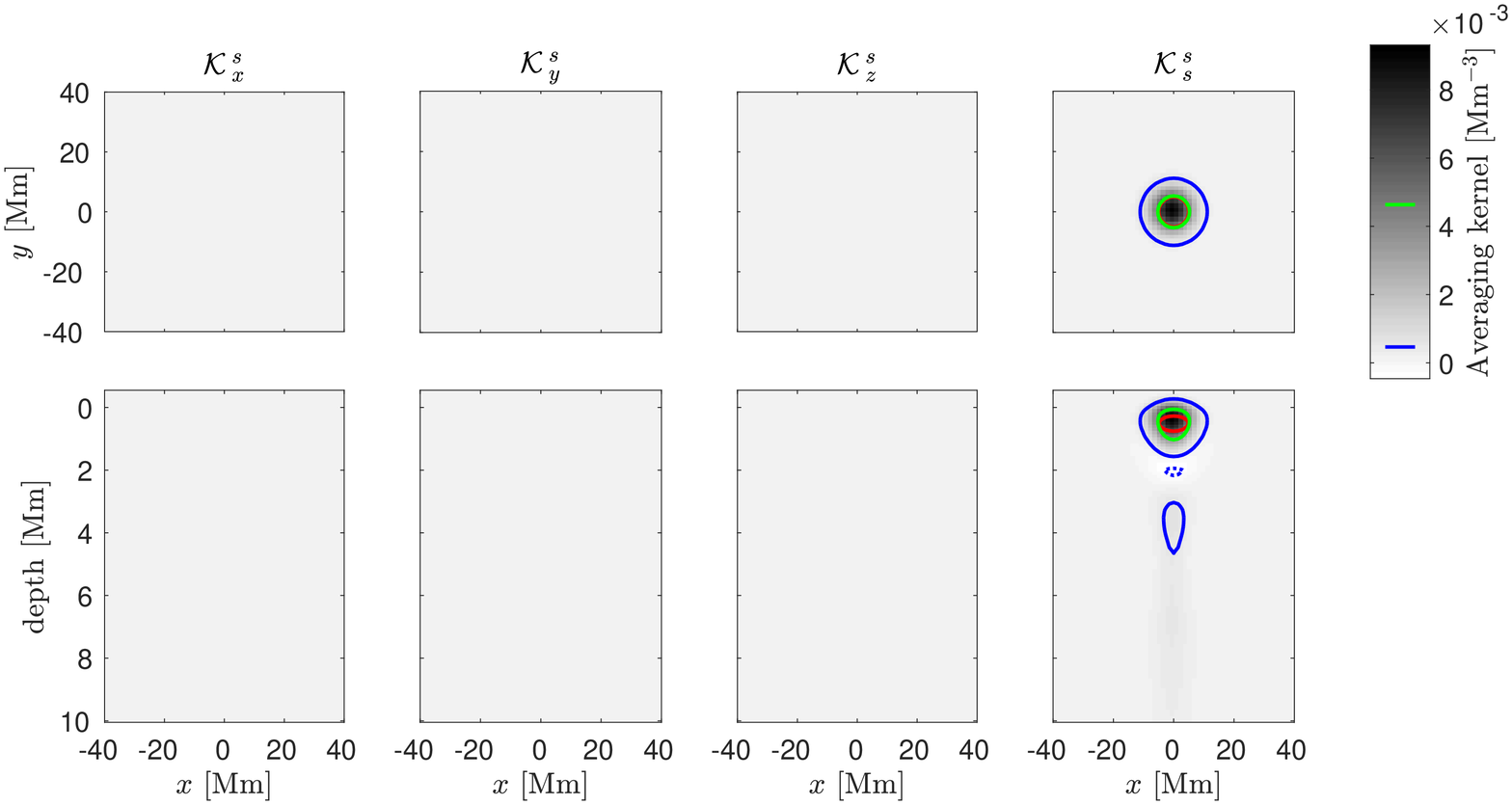}
{Averaging kernel for the inversion for the \ssp{} at the target depth of 0.5~Mm (positive $\dcs$ at the supergranule centre). See the caption of Fig.~\ref{fig:vx_0.5_rakern} for details. }
{fig:cs_0.5_rakern_ok}

\wfigure{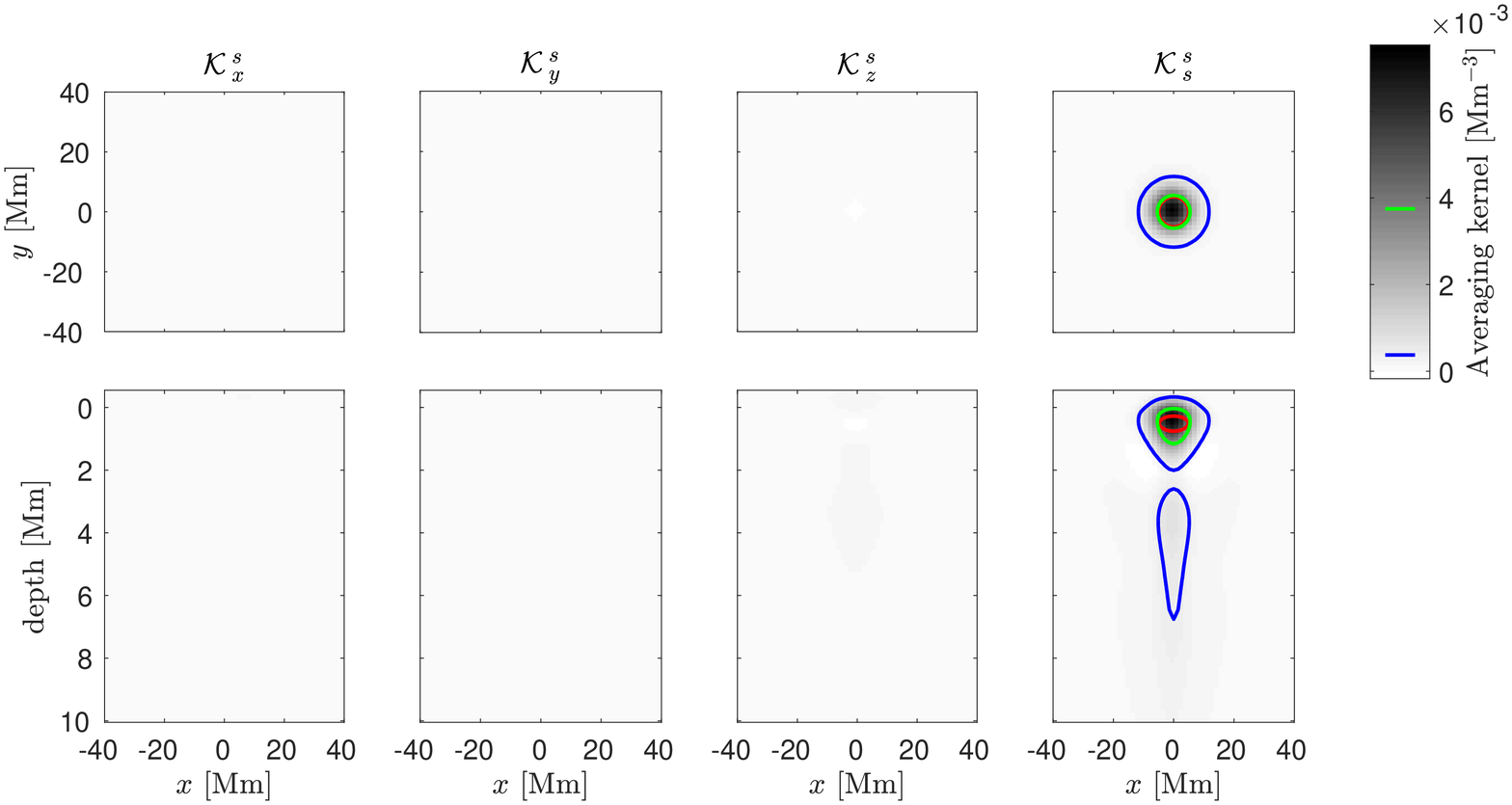}
{Averaging kernel for the inversion for the \ssp{} at the target depth of 0.5~Mm (negative $\dcs$ at the supergranule centre). See the caption of Fig.~\ref{fig:vx_0.5_rakern} for details. }
{fig:cs_0.5_rakern_ko}

Two quantities entered our inversions, the noise covariance matrices $\lab$ and the sensitivity kernels $K$, thus logically, the issue should be identified in one of these. First, let us assume the problematic input quantities were the noise matrices. As the noise matrices were computed from the measured travel times, we should have found issues in the travel times. 

There might be two kinds of issues in the measured travel times. First, the reference cross-correlation function was computed as the long-term average cross-correlation. This cross-correlation was not fully consistent with the reference Model~S used in the computation of the sensitivity kernels. To investigate this possibility, we computed another set of the travel times which were consistent with the reference model. The differences between both sets of the travel times were negligible. 

Second, the travel times contained a systematic centre-to-limb trend caused by foreshortening. Centre-to-limb variations of the travel-time differences have been recorded in the past \citep[e.g.][]{Zhao2012}. These recorded variations were on the order of 1--2 seconds and their origin is not in the foreshortening. \citet{Baldner2012} proposed that part of this effect may be explained by the asymmetrical nature of solar granules which affect the phase shift of the  modes as a function of the height and thus the distance to the disc centre. Our systematic centre-to-limb effects are caused purely by the plain-parallel approximation, and its magnitude is much larger than that of \citet{Zhao2012}.

Moreover, the foreshortening trend seen in our travel times was identical in the positive and negative travel times, hence it did not affect the difference travel-time geometries (o-i, e-w, n-s) where it was  removed by the subtraction. On the other hand, the trend was preserved in the mean travel times. The procedure of the foreshortening subtraction is described in Appendix~\ref{app:trend}.

\tcfigure{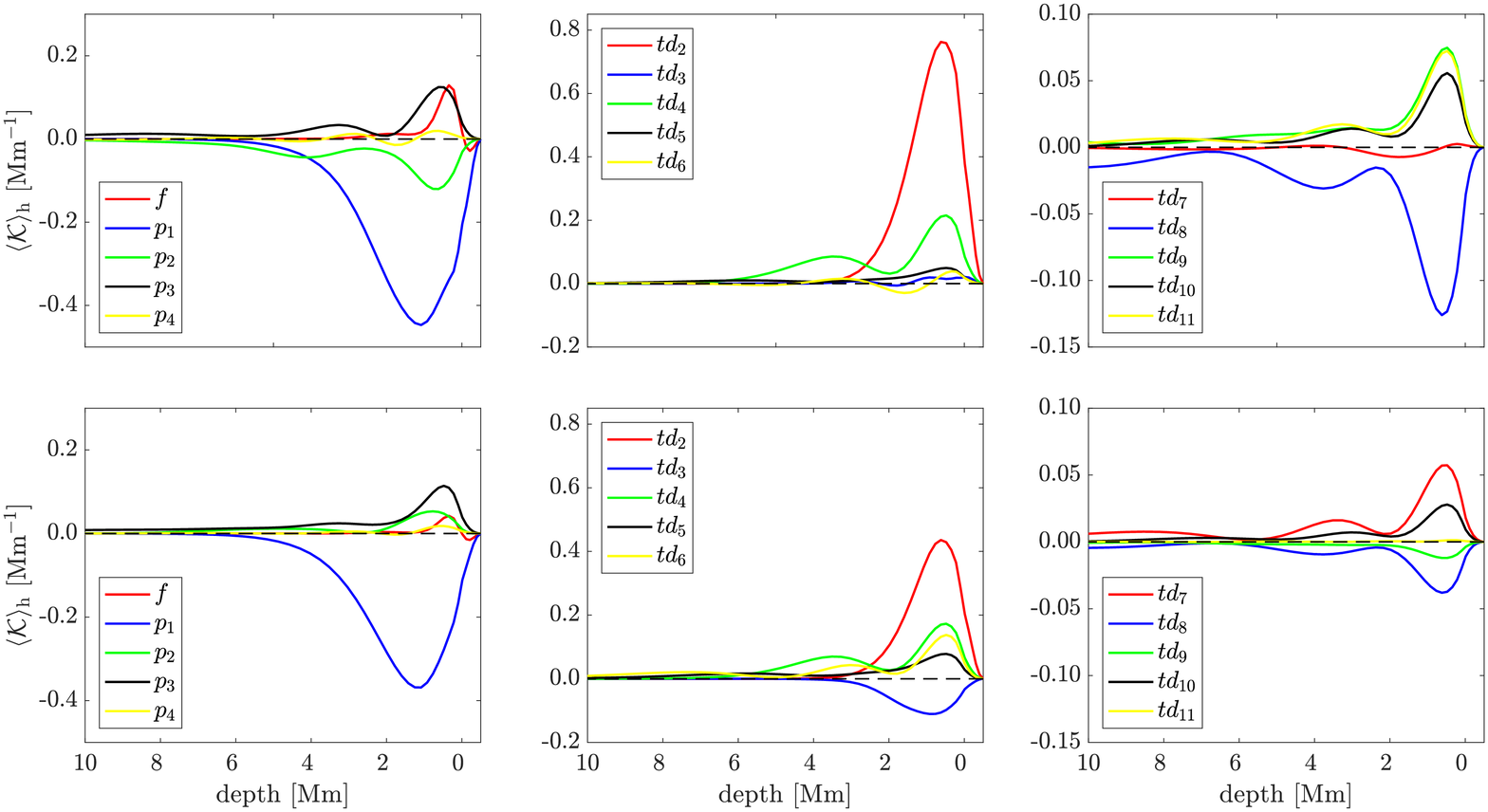}
{Contributions to the horizontally averaged averaging kernel for the \ssp{} inversions at 0.5~Mm depth. Top row: Inversion with the positive $\dcs$ at the supergranule centre. Bottom row: Inversion with the negative $\dcs$ at the supergranule centre. Left column: Ridge filters. Middle column: Phase-speed filters from $td_2$ to $td_6$. Right column: Phase-speed filters from $td_7$ to $td_{11}$. }
{fig:cs_0.5_rakern_h}

The problematic inverted quantities, the vertical flows, and the \ssp{} are sensitive to the mean travel times. Therefore, they are sensitive to the method of trend subtraction. On the other hand, \citet{Svanda_2015} found similar issues in all the components of the vector flows without considering the mean travel times. It would therefore seem unlikely that the trend in travel times and consequently in the noise matrices should be responsible for the fragile results of the inversions. Additionally, our inversions were dominated by the misfit term, because the noise was suppressed by the ensemble averaging. The issues remained even when the $\mu$ parameter, which controlled the noise regularisation, was very small and the noise matrices became less important. Therefore, the noise matrices were likely not the source of the issues.

The other suspicious input quantities are the sensitivity kernels, which contribute both to the inversion and to the averaging kernels. The sensitivity kernels result from the forward modelling, where the disagreement between the real Sun and the model would result in the model error, which is unconstraint in our inversion. \citet{Boning_2016} compared two different models of the sensitivity kernels. They found small discrepancies of about 0.3\% in the integrals of the corresponding sensitivity kernels. Larger differences were discovered in the detailed structure of the corresponding kernels. In our inversions, we combined 520 sensitivity kernels in order to reduce the misfit term. The combination of many sensitivity kernels could amplify the systematic errors hidden in our sensitivity kernels.
\tcfigure{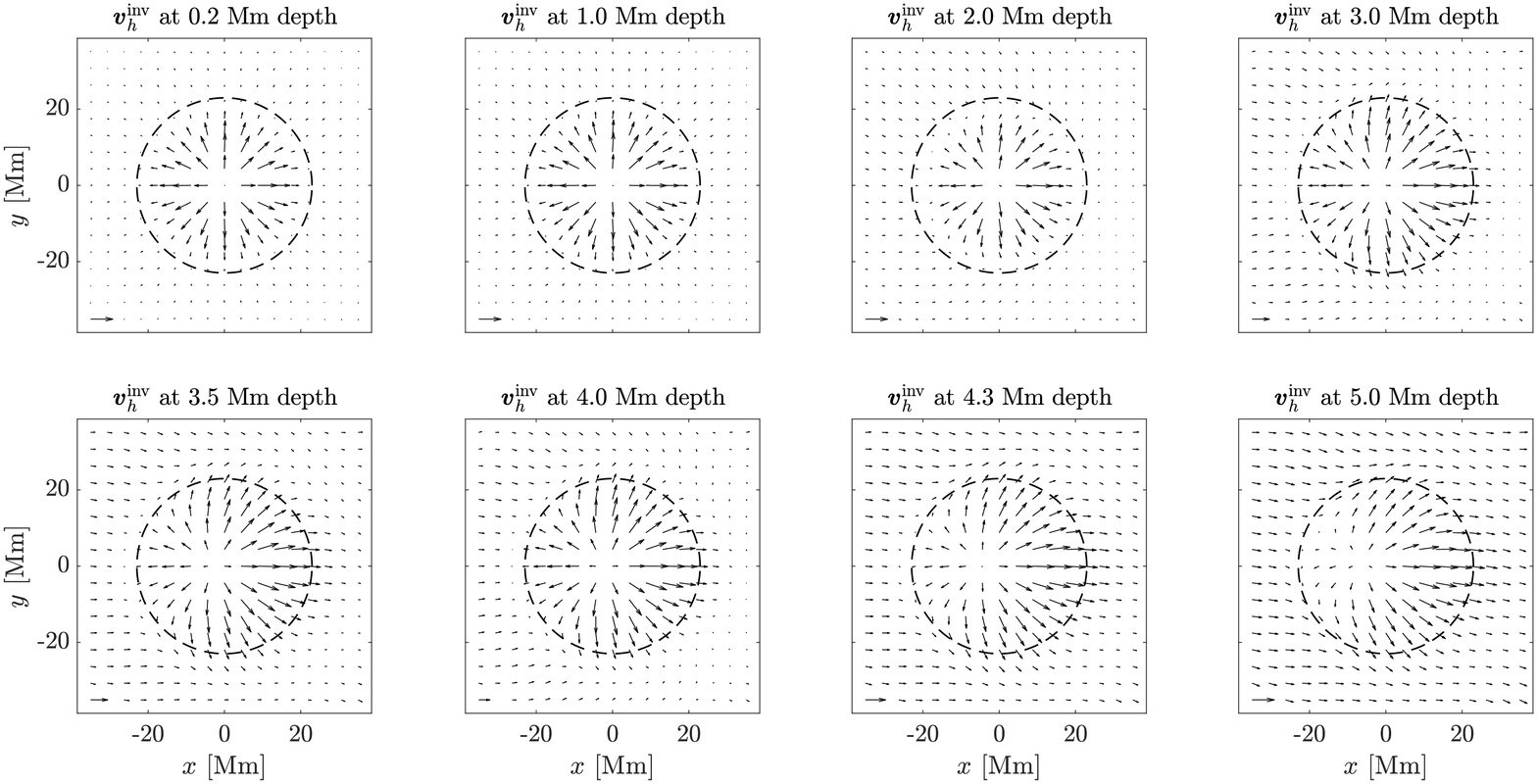}
{Inverted horizontal flows. The reference arrows in the bottom left-hand corners correspond to 250~\mps{} (0.2 and 1.0~Mm), 100~\mps{} (2.0 and 3.0~Mm), and 50~\mps{} (3.5, 4.0, 4.3, and 5.0~Mm). At depths of 3.0~Mm and greater, one can see the systematic longitudinal flows. The radii of the black dashed circles are approximately 23~Mm. }
{fig:vh_all}

To investigate this possibility, we compared the individual mode contributions to the averaging kernels, which are plotted in Fig.~\ref{fig:cs_0.5_rakern_h}. By comparing the contributions of the individual modes to the two inversions resulting in the $\dcs$ inversion with the opposite sign, we see that the contributions change. In some cases, the contributing modes even flipped the sign (e.g. $p_2$ mode in the left panel, $td_3$ mode in the middle panel, and $td_9$ mode in the right panel). Should the sensitivity kernels for these modes be unrealistic,
the change of sign of the inverted quantity could be explained.

Even though the inverted vertical flows seem to be inconsistent, the surface inversion models based  solely on the $f$-mode travel times are comparable with other measurements, for example the observed Dopplergrams. \citet{SvandaRoudier_2013} performed the $f$-mode surface inversions for the vector flows with the difference averaging geometries only. Therefore, the authors inverted for the horizontal variations of the vertical flows. They found excellent agreement between the inverted horizontal flows and proper motion of granules (see their Fig.~2). Furthermore, they compared the observed Dopplergram and the inverted vector flows projected onto the line of sight. Although they did not invert for the full vertical flows, the agreement was convincing (see their Fig.~5). 

We note that the $f$-mode travel times are not sensitive to the \ssp{} \citep{Burston_2015}, and so it is impractical to use a similar setup to investigate \ssp{} at the surface. 


\subsection{Shallow inversions for vector flows}
\label{sec:shallow_flows}

We computed the first set of inversions at the target depths of 0.2, 0.5, 1.0, 2.0, 3.0, 3.5, 4.0, 4.3 and 5.0~Mm (the parameters of the target functions are given in Table~\ref{tab:target_1}). At depths smaller than 2.0~Mm the inverted vertical flows can be either upflows or downflows depending on the trade-off parameters, even though the averaging kernels of the two different results were comparable. At depths equal to or greater than 2.0~Mm, the inverted vertical flows are always downflows with a large amplitude (which is not consistent with the continuity equation) regardless of the combination of the trade-off parameters. However, at depths greater than 3~Mm, the inversions for the vertical flows are dominated by the regularisation terms. Therefore, the inversions tend to shallow inversions. The same holds true for the \ssp{}. The only differences are the localisations of the inversions. The inversions for the \ssp{} have reasonable localisations down to a depth of about 4~Mm.

\begin{table}
    \caption{Parameters of the 3D Gaussian target functions used in Sect.~\ref{sec:shallow_flows}. The table gives the target depth $z_0$ and the horizontal $\FWHM{}_h$ and vertical $\FWHM{}_z$ extents of the 3D Gaussian in the form of the full width at half maximum.}
    \label{tab:target_1}
    \centering
    \begin{tabular}{c c c}
        \hline\hline
        $z_0$ & $\FWHM{}_h$ & $\FWHM{}_z$\\
        \hline
        0.2~Mm & 10~Mm & 3.2~Mm\\
        0.5~Mm & 10~Mm & 0.5~Mm\\
        1.0~Mm & 10~Mm & 0.5~Mm\\
        2.0~Mm & 10~Mm & 1.0~Mm\\
        3.0~Mm & 10~Mm & 1.0~Mm\\
        3.5~Mm & 15~Mm & 2.5~Mm\\
        4.0~Mm & 10~Mm & 1.0~Mm\\
        4.3~Mm & 15~Mm & 1.5~Mm\\
        5.0~Mm & 10~Mm & 1.0~Mm\\
        \hline
    \end{tabular}
\end{table}

The inverted horizontal flows are consistent at all the target depths. At depths of about 2~Mm and shallower, the inverted horizontal flows are purely divergent. At greater depths, we detect systematic flow in the longitudinal direction to the west. Therefore, the divergent flows change to more laminar in this direction. In Fig.~\ref{fig:vh_all} we plot the inverted horizontal flows at all the target depths except for the inversion at 0.5~Mm which is shown in Fig.~\ref{fig:flows_0.5}. The amplitude of the inverted horizontal flows changes with the depth. Therefore, the reference arrows at the left bottom corners differ. At the depths 0.2 and 1.0~Mm, the reference arrows correspond to 250~\mps{}, at depths of 2.0 and 3.0~Mm they correspond to 100~\mps{}, and at the depths of 3.5~Mm and greater the arrows correspond to 50~\mps{}. The systematic longitudinal flow is clearly visible at depths of 3.0~Mm and greater. We discuss the systematic flow in Sect.~\ref{sec:vx_vs_rot}. At all depths, the divergent character of the latitudinal flows is evident. No significant systematic flow is detected in the latitudinal direction because the averaging process naturally mixes up both hemispheres. 

\tcfigure{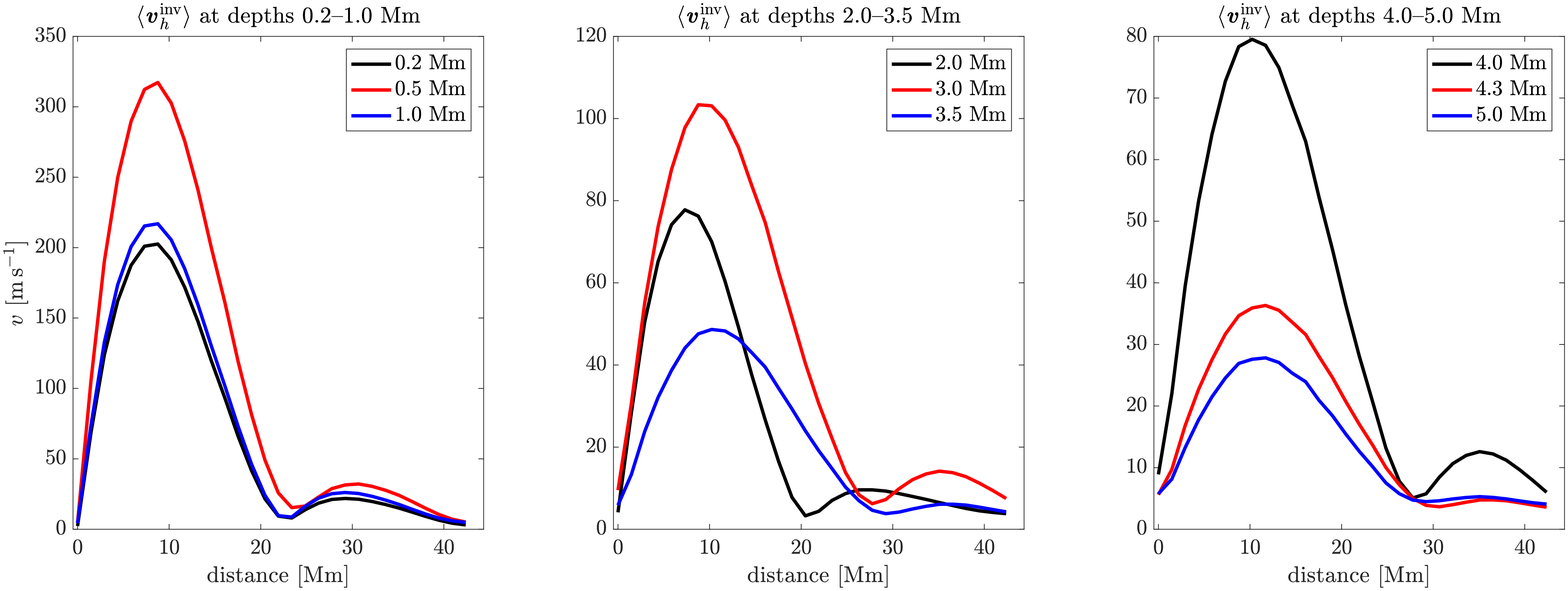}
{Average outflows as a function of distance from the centre of the average supergranule and depth. The local minima at around 23~Mm roughly correspond to the edges of the average supergranule as a function of depth. We note that the vertical axes differ. }
{fig:vh_dist}

In Table~\ref{tab:vh} we summarise the maximum and the mean horizontal outflows at all the target depths. We computed the maximum outflows as an average of maximum outflows in the given directions (longitudinal flows or latitudinal flows). This approach naturally subtracts the systematic flow. As seen in Table~\ref{tab:vh}, the amplitudes of the horizontal flow components are comparable after the systematic flow is removed. The mean outflows were computed as the maximum of the mean outflows over an annulus as a function of the distance from the supergranular centre (maximum values in Fig.~\ref{fig:vh_dist}). Here, the systematic flow was subtracted. This naturally covers possible changes in the size of the average supergranule with depth. The annulus width is about 1.4~Mm. The mean outflows should have been the robust estimate because it was the average outflows over the whole polar angle. The average outflows as a function of the distance from the centre of the average supergranule and the depth are plotted in Fig.~\ref{fig:vh_dist}.

The non-monotonic horizontal outflows are caused by the depth localisations of the inversions (see Fig.~\ref{fig:akerns_v1}). The inversion at 0.2~Mm has significant sensitivity down to 2-Mm depth which lowers the amplitude. The inversions at 0.5 and 1.0~Mm are well-localised. The averaging kernels of the 2Mm inversion have a negative contribution at depths where the high-amplitude inversion has maximum sensitivity. This again might decrease the inverted amplitude. At 3.0~Mm the negative contribution is smaller and shifted to greater depths. Therefore, the effect on the amplitude is smaller too. The inversion at 3.5~Mm has maximum sensitivity around the targeted depth, but the contribution from depths where the horizontal flows are small is significant. The other averaging kernels have similar properties, which correspond to the continuous decrease in amplitude of the horizontal flows at these depths. This demonstrates the requirement for knowledge of the averaging kernels. We note that the side lobes in the vertical direction are consequences of a fixed horizontal extent of the target functions.

\begin{table}
    \caption{Maximum and mean horizontal outflows as a function of depth in the average supergranule. The systematic flow was subtracted.}
    \label{tab:vh}
    \centering
    \begin{tabular}{c c r r r}
        \hline\hline
        $z_0$ & $\langle z \rangle $ & $\max \left( v^{{\rm inv}}_x \right)$ & $\max \left( v^{{\rm inv}}_y \right)$ & $\max \left(\left< \vec{v}^{{\rm inv}}_h \right> \right)$\\
        \hline
        0.2~Mm & 1.0~Mm & 209~\mps{} & 207~\mps{} & 203~\mps{}\\
        0.5~Mm & 0.8~Mm & 328~\mps{} & 324~\mps{} & 317~\mps{}\\
        1.0~Mm & 1.3~Mm & 222~\mps{} & 221~\mps{} & 217~\mps{}\\
        2.0~Mm & 2.6~Mm & 78~\mps{} & 79~\mps{} & 78~\mps{}\\
        3.0~Mm & 3.4~Mm & 108~\mps{} & 108~\mps{} & 103~\mps{}\\
        3.5~Mm & 3.9~Mm & 51~\mps{} & 51~\mps{} & 49~\mps{}\\
        4.0~Mm & 3.9~Mm & 85~\mps{} & 84~\mps{} & 80~\mps{}\\
        4.3~Mm & 4.7~Mm & 39~\mps{} & 38~\mps{} & 36~\mps{}\\
        5.0~Mm & 5.7~Mm & 31~\mps{} & 30~\mps{} & 28~\mps{}\\
        \hline
    \end{tabular}
\end{table}

\ocfigure{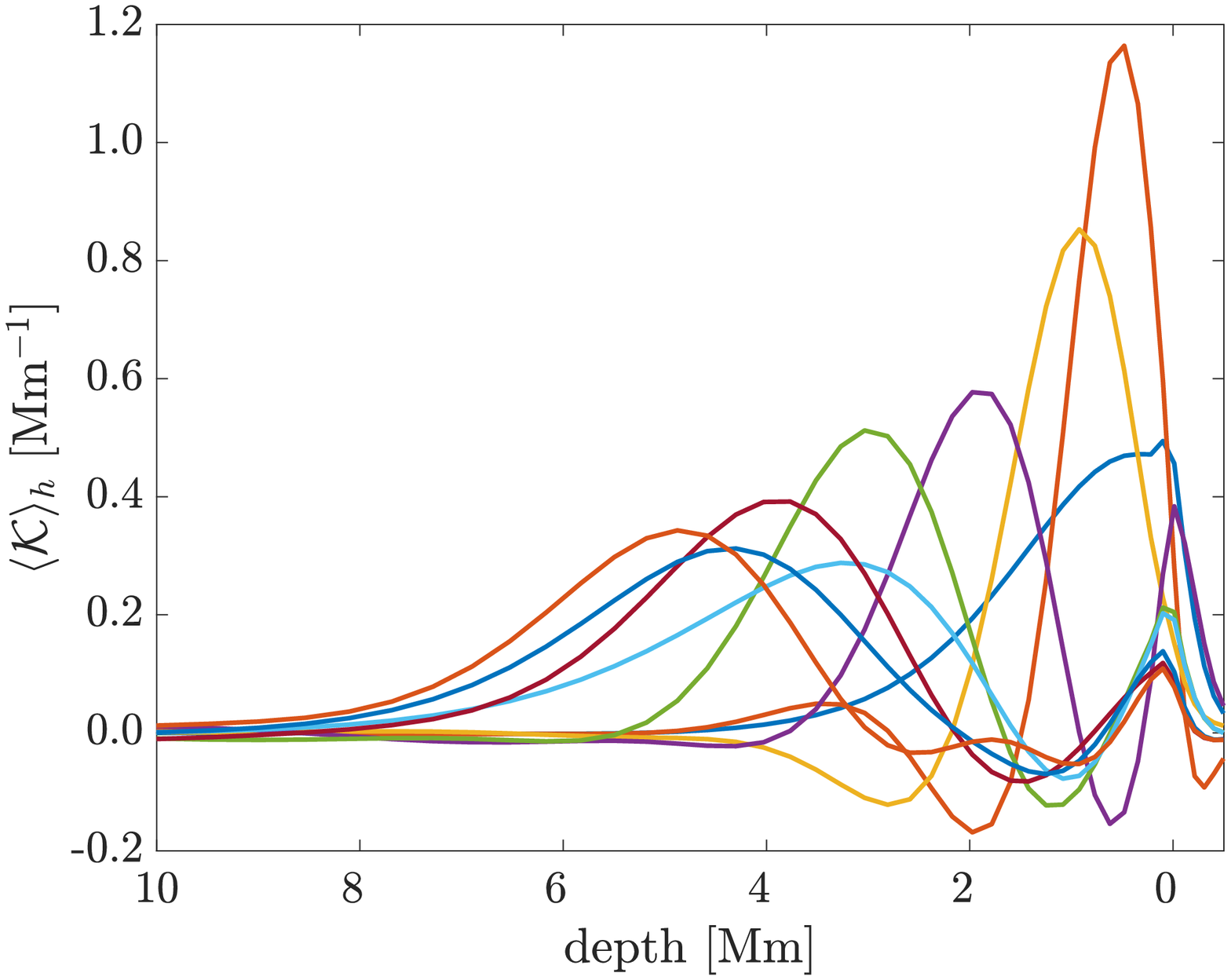}
{Horizontally averaged averaging kernels for various inversions for the horizontal flows. The inversion models are well-localised for all the target depths.}
{fig:akerns_v1}


\subsection{Deep inversions for vector flows}
\label{sec:deep_flows}

In the previous section, we show that the inversions for the horizontal flows provide us with reasonable flow estimates, but the inverted vertical flows are trustworthy only in the case of the surface inversion. Therefore, in the following, we focus on the horizontal flows only. We are interested in the depth structure of the average supergranule. For this reason, we performed another set of inversions for the horizontal flows from the surface down to the depth of 25~Mm.  

To gain a complete view of the vector flows in the average supergranule, we computed the estimate of the vertical flows $v^{{\rm est}}_z$ from the horizontal flows by integrating the continuity equation from the surface:
\begin{equation}
    v^{{\rm est}}_z \left( \vec{r}_0; z_0 \right) = \frac{-1}{\rho \left(z_0 \right)} \int \limits_0^{z_0} \dif{z} \nabla_h \cdot \left[\rho \left(z \right)\, \vec{v}^{{\rm inv}}_h \left(\vec{r}_0, z \right) \right],
\end{equation}
where $\rho$ is the density and $\nabla_h$ the horizontal gradient. We note that the same methodology was used by \citet{Greer_2016}. The authors also introduced a set of target functions (see their Appendix A) which we adopted too. The parameters of the target functions are summarised in Table~\ref{tab:target_2}. Additionally, we completed this set with a surface $f$-mode-only inversion which was validated against the independent measurements by \cite{SvandaRoudier_2013}. At the surface, we were also able to include the inverted vertical velocity. 

\begin{table}
    \caption{Parameters of the 3-D Gaussian target functions utilised in Sec.~\ref{sec:deep_flows}.}
    \label{tab:target_2}
    \centering
    \begin{tabular}{c c c}
        \hline\hline
        $z_0$ & $\FWHM{}_h$ & $\FWHM{}_z$\\
        \hline
        1.0~Mm & 10~Mm & 1.0~Mm\\
        3.0~Mm & 14~Mm & 1.0~Mm\\
        5.0~Mm & 18~Mm & 1.2~Mm\\
        7.0~Mm & 22~Mm & 1.5~Mm\\
        9.0~Mm & 26~Mm & 1.8~Mm\\
        11.0~Mm & 30~Mm & 2.3~Mm\\
        13.0~Mm & 34~Mm & 2.9~Mm\\
        15.0~Mm & 38~Mm & 3.6~Mm\\
        17.0~Mm & 42~Mm & 4.4~Mm\\
        19.0~Mm & 46~Mm & 5.3~Mm\\
        21.0~Mm & 50~Mm & 6.3~Mm\\
        23.0~Mm & 54~Mm & 7.5~Mm\\
        25.0~Mm & 58~Mm & 8.7~Mm\\
        \hline
    \end{tabular}
\end{table}

The corresponding horizontally averaged averaging kernels are plotted in Fig.~\ref{fig:akerns_v2}. As seen, the inverse models are reasonably localised down to the depth of about 17~Mm, and the inversions with target depths greater than 17~Mm have peaks a few Mm shallower than expected. 

\ocfigure{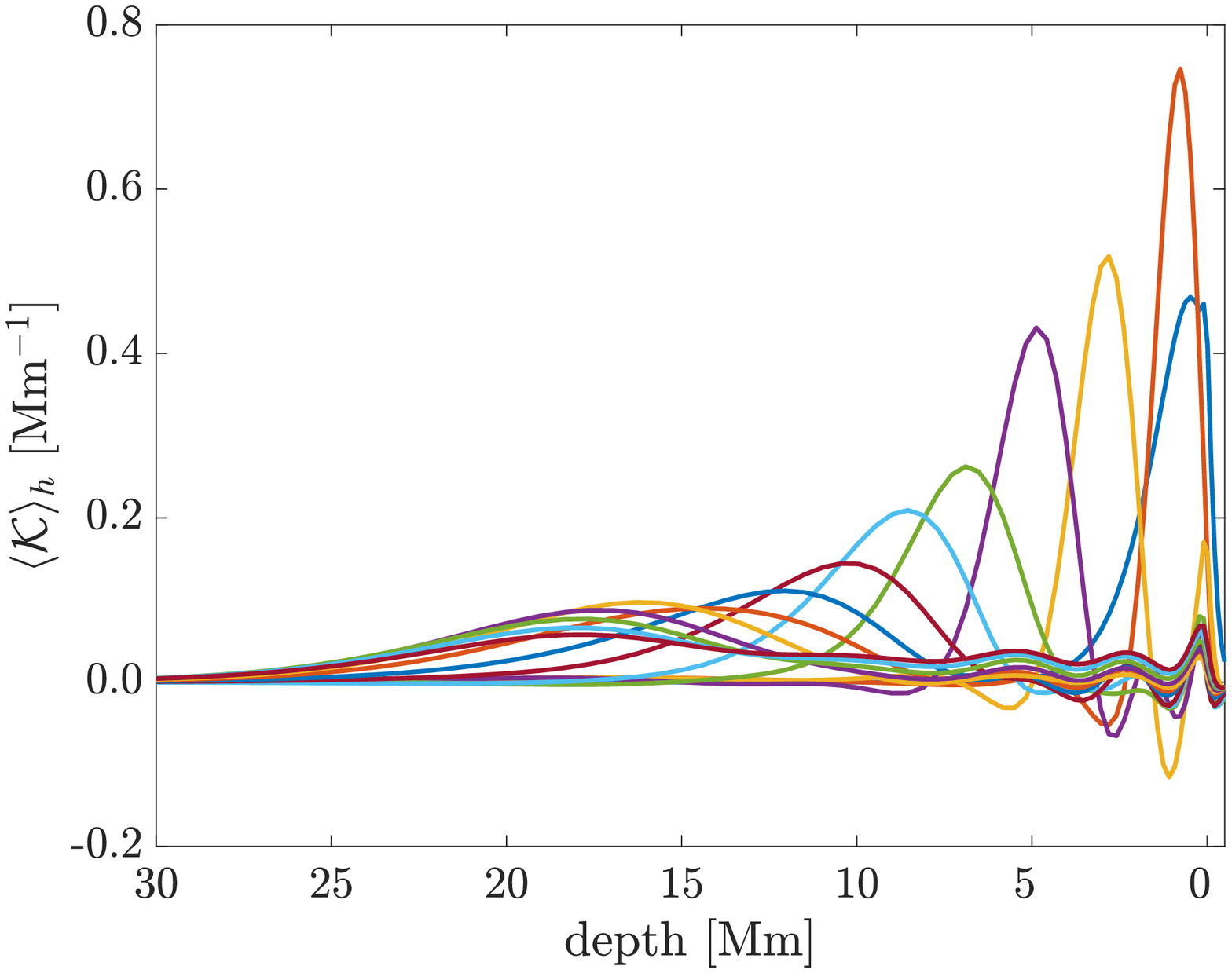}
{Horizontally averaged averaging kernels for various inversions for the horizontal flows. The inversion models were well-localised down to the depth of about 17~Mm. }
{fig:akerns_v2}

The inverted horizontal flows were purely divergent (after the systematic flow was subtracted) down to a depth of about 6--7~Mm. At greater depths, the amplitude of the horizontal flows is small, and the divergent flows become slightly convergent. When plotting the non-averaged snapshots of the flows in the $(x\text{-}z)$ or $(y\text{-}z)$ planes, we noticed that the coherent structures corresponding to supergranules are not purely vertical in these plots, and therefore these structures are not continuous in a purely vertical direction. Instead, they depict lateral displacements with depth. In our average-supergranule ensemble averaging scheme, the neighbouring structures therefore might mix and affect the derived flows. Similar structures can also be seen in Fig.~4 of \citet{Greer_2016} for example.

In Fig.~\ref{fig:v_cuts} we plot the vertical cuts of the inverted horizontal flows (left and middle panels) and the computed vertical flows (right panel). It would seem that the supergranules (at least the strong ones preferred by our detection routine) are coherent structures going quite deep down to 15~Mm and possibly deeper (see the cut in Fig.~\ref{fig:vz_centre_cut} for a better visibility). This is in part because the $\FWHM_h$ of deeper inversions are comparable to or greater than the diameter of the average supergranule at the surface and also because of the lateral depth displacement described in the previous paragraph. The horizontal flows are strongest near the surface and their amplitude decreases monotonously with depth. The horizontal flow seems to reach the amplitude comparable to the background at the depth of about 7~Mm, where an apparent horizontal flow `reversal'  also appears. On the other hand, the horizontal flow magnitude of this reversed horizontal flow remains comparable to the background, and therefore it is not possible to consider the detection of this reversal as significant. 

The vertical flow in the centre of the average supergranule has a surface upflow magnitude of about 5~\mps{} that was inverted from the $f$-mode travel times. Our estimates based on the integration of the continuity equation therefore indicate that the vertical upflow magnitude first increases with depth and reaches the maximum of about 35~\mps{} at a depth of about 3~Mm (see Fig.~\ref{fig:vz_centre_cut}). After this peak, the magnitude of the vertical flow further decreases with depth. The vertical flow remains positive (an upflow) within the range of the studied depths. Our results seem consistent with the findings of \citet{Greer_2016}. We note that the transition between the surface (inverted/measured) and the depths of 1--3~Mm (integrated from the continuity equation) is continuous, which strengthens the validity of both approaches. 

\tcfigure{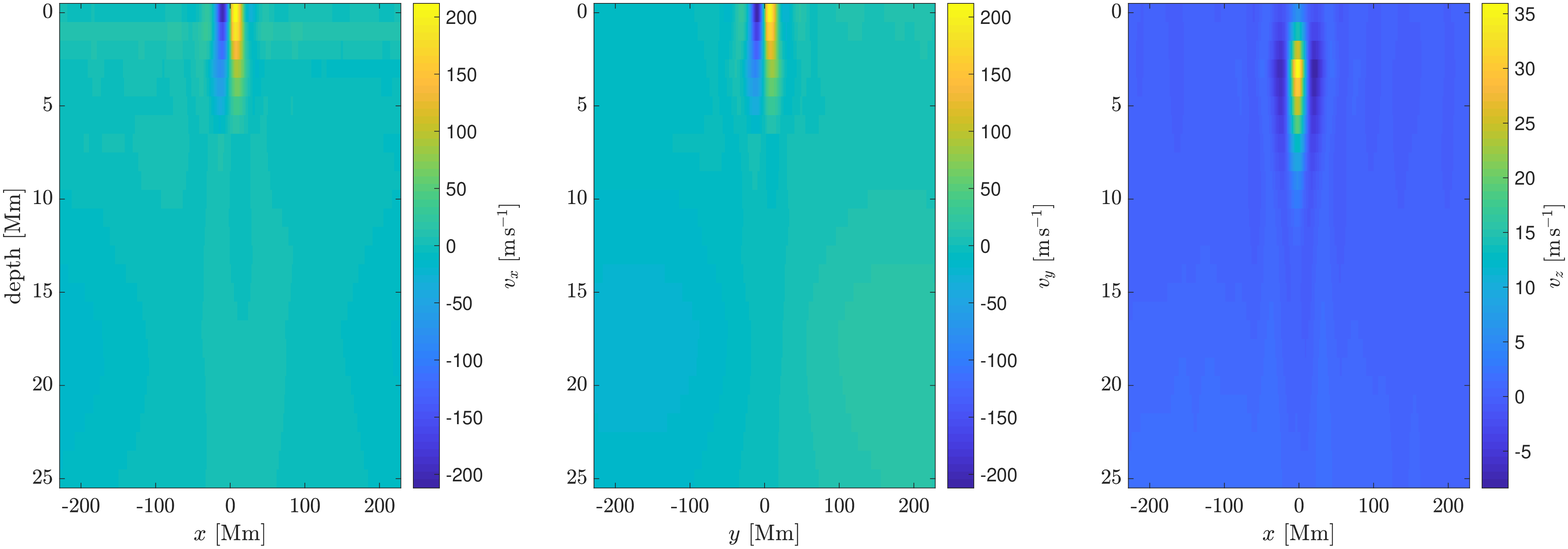}
{Vertical cuts through the inversion models. Left: Longitudinal flows without the systematic flow. Middle: Latitudinal flows. Right: Vertical flows. The colour bars of the horizontal flows are the same. }
{fig:v_cuts}

\ocfigure{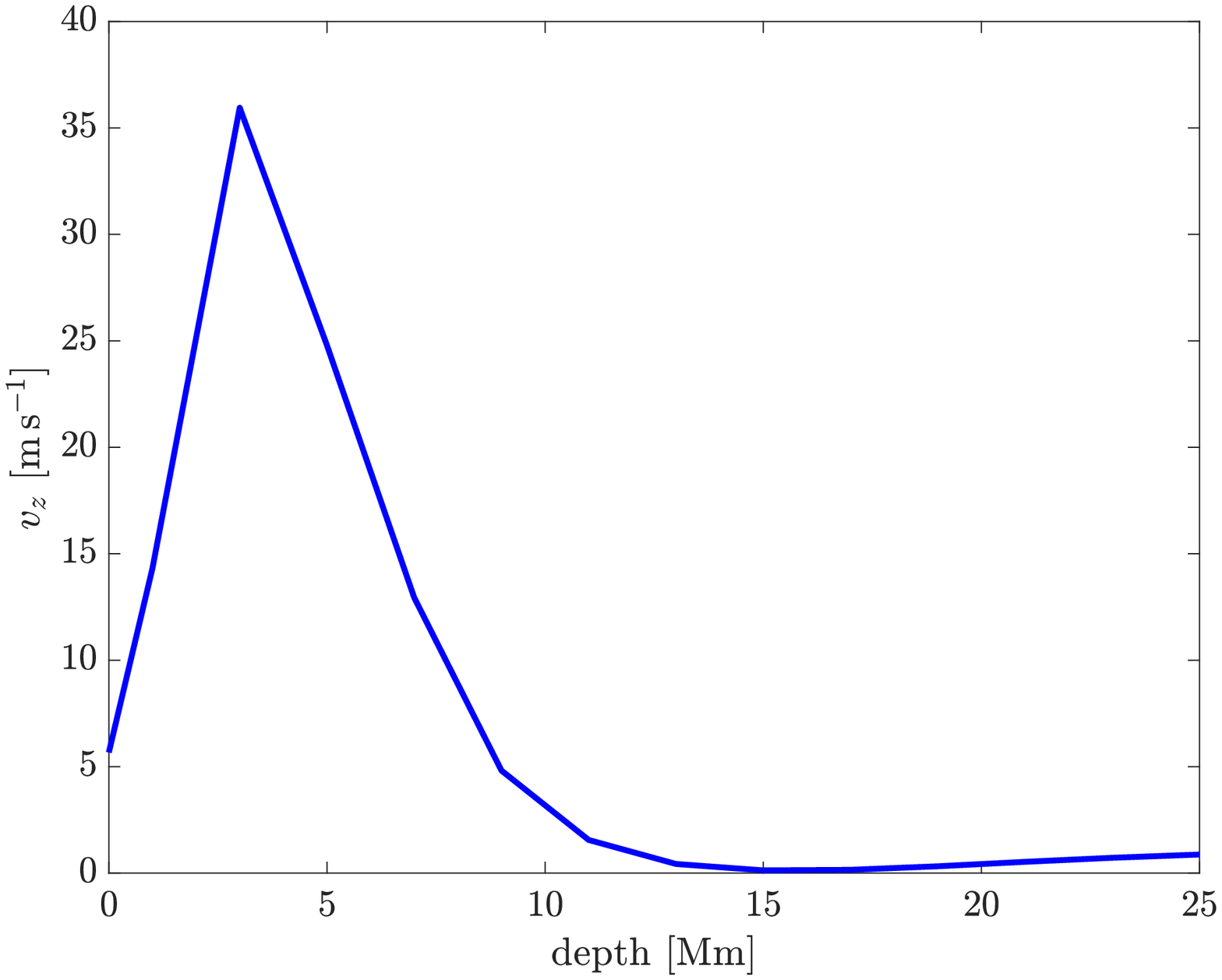}
{Vertical profile of the vertical flows at the centre of the average supergranule. }
{fig:vz_centre_cut}


\subsection{Systematic flow in the longitudinal direction and its relation to solar rotation}
\label{sec:vx_vs_rot} 

As seen in Fig.~\ref{fig:vh_all}, there is a systematic component in the longitudinal flows. We analysed the systematic flow at all depths and compared it with the changes in rotation rate at the corresponding depths. To construct the depth dependence of the systematic flow,  we first filtered out edges and central parts of the inverted maps of the longitudinal flows. The radii of the filtered central areas were about two and half diameters of the average supergranule. This suppresses possible inhomogeneities in the horizontal outflows connected with flows inside supergranules. 

To compare with the solar rotation, we used the solar rotational rate model of the equatorial regions published by \citet{Howe_2000}. From this model, we subtracted the surface rotation rate because our inverted results were in the Carrington-rotation rest frame. To account for the depth localisations of our inversions, we smeared the model of the rotation rate with the set of the averaging kernels. We reiterate that for this study we computed two different sets of inversions with different averaging kernels: one set with a denser sampling for the near-surface layers, and another set with a scarcer depth sampling going down to the depth of 25~Mm. 

The mean systematic flows and the rotation rates for all the target depths are visualised in Fig.~\ref{fig:vx_trend}. In the main plot there are results for the second set, and in the subplot there are results of the first (shallow) set. The systematic flows and the rotation rates are comparable from the surface down to the deepest inversion at 25~Mm. The correlation coefficients of the two curves and the typical difference between the systematic flow and the model of rotation rate were 0.95 and 0.96 and 5~\mps{} and 10~\mps{} for the first and the second set, respectively. We note that the decrease in rotation rate at the first Mm of the set 1 is due to the depth localisations of our inversions. The discrepancy between the inverted systematic longitudinal flows and the rotation model might be related to the fact that the averaged supergranules were not located only at the equator but also at higher latitudes. The rotation rate is lower at higher latitudes. According to the model of \citet{Howe_2000}, the rotation speed at the latitude of 20$^\circ$ (we looked for supergranules up to about this latitude) is about 30~\mps{} slower than the rotation speed at the equator. The detection algorithm looks for supergranules in a circular area with the centre at the equator (see middle panel of Fig.~\ref{fig:ave_sg}). Hence, this approach partially suppresses this effect. Therefore, the inverted systematic longitudinal flows were naturally averaged over a few degrees in latitude, and therefore the systematic flows must have been smaller. We note that this effect is not covered in the plotted error bars. The indicated error bars are based on the uncertainty estimates of the inversions scaled by the $\sqrt{N}$ factor, where $N$ is the number of considered supergranules. The systematic flows were smaller than the rotation model down to 13~Mm below the surface. The systematic flows are larger than the rotation model at depths of 13~Mm and greater, the difference between the two is up to 15~\mps. 

In general, we cannot expect a perfect match for several reasons. First of all, the methodologies behind the two data series are very different. The methods are the local helioseismology in a plain-parallel model in our case and the spherical global helioseismology in the case of the considered model of rotation. Second of all, both trends in zonal velocity were obtained from observables secured in different time periods, separated by many years. The solar rotation is known to change with time.

The goal of this exercise is to show that the depth-dependence of the solar rotation measured as a by-product of our supergranular-scale flow inversions resembles the model of rotation published elsewhere. We consider this resemblance as significant proof that our inversions for horizontal flows perform well. 

\ocfigure{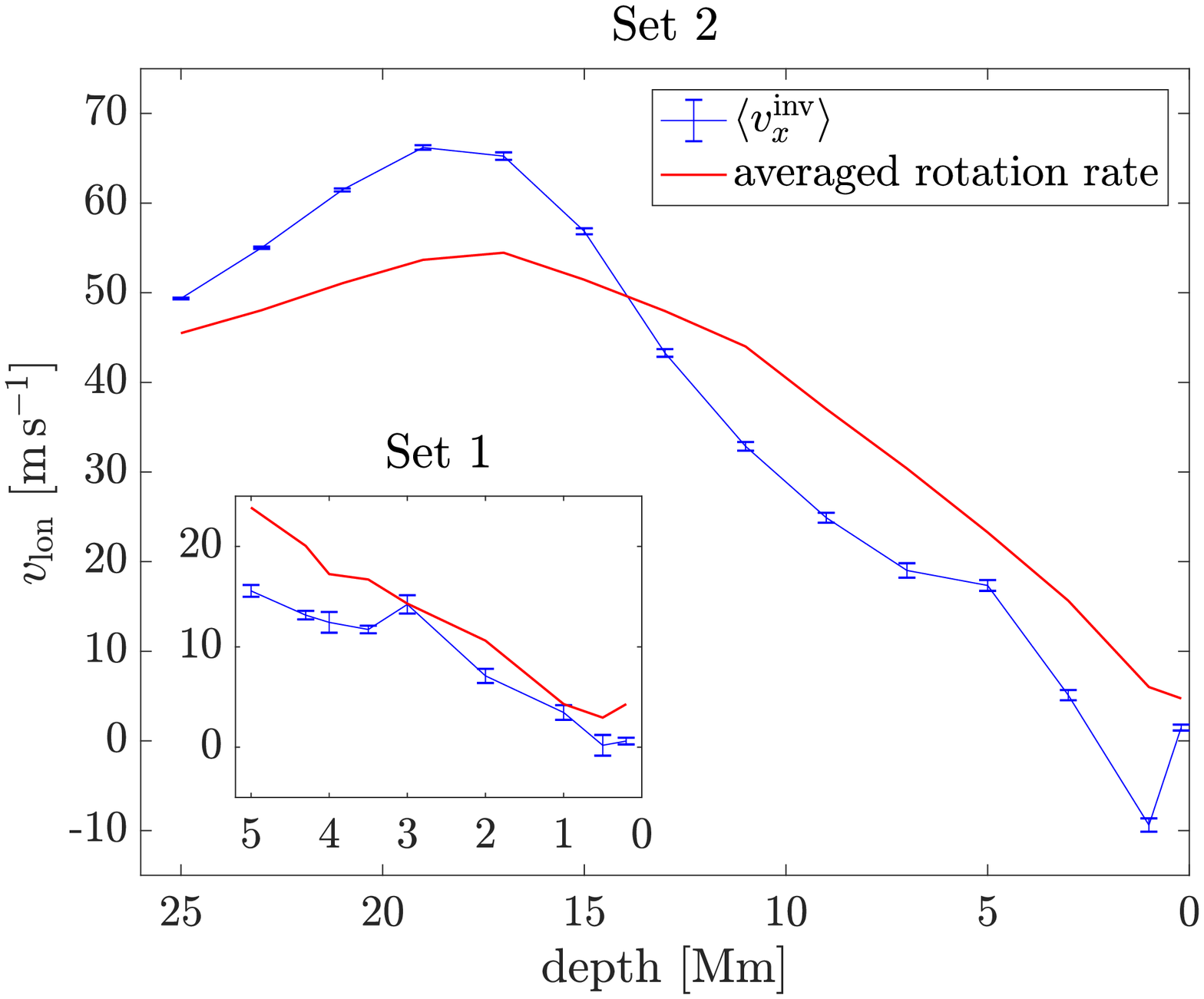}
{Mean systematic flow in the longitudinal direction and the smeared model of the rotation rate. }
{fig:vx_trend}


\section{Conclusions}

We performed two sets of the helioseismic inversions for the vector flows and the \ssp{} and applied the ensemble averaging methodology in order to learn about the structure of the average supergranule. To this aim, we used the \mbox{MC-SOLA} method. The selected target depths were 0.2, 0.5, 1.0, 2.0, 3.0, 3.5, 4.0, 4.3, and 5.0~Mm in the first set and from 1~Mm to 25~Mm with incremental sampling every 2~Mm in depth completed with the surface inversion in the second set.

The noise suppression does not significantly improve the localisations of the vertical-flow inversions. The reasonable localisations were achieved down to a depth of about 3~Mm. Moreover, we find similar issues which have also been discussed by \citet{Svanda_2015}, although our inversion setup is different. Even though the averaging kernels are comparable, the directions of the inverted vertical flows are opposite. The direction of the vertical flows often changes with the depth as well. The changes of the vertical-flow directions might be related with the cross-talk. Highly symmetric horizontal flows with large amplitude could easily spoil the inverted vertical flows. However, we do not find any differences even in the cases of very strong regularisation of the cross-talk. We also tested the input quantities, that is, the sensitivity kernels and the travel times. We do not find any obvious suspicious behaviour for either of them. The sources of the issues are probably related to the completely unknown systematic errors of the sensitivity kernels. These systematic errors might be amplified as the number of sensitivity kernels in the inversion increases.

The inversions for the \ssp{} have similar properties. The inverted results could be either positive (increase in sound speed) or negative (decrease in sound speed), while the corresponding averaging kernels are comparable. These issues with the \ssp{} were not published by other authors. The noise suppression improves the localisations of the inversions. The averaging kernels are localised around the target depth down to the depth of about 4~Mm.

The inverted horizontal flows are consistent with themselves. We did not meet any issues with the changes in the flow directions at any depth. We find a divergent flow at all depths. The inverted horizontal flows from the first set of inversions are plotted in Figs.~\ref{fig:flows_0.5} and \ref{fig:vh_all}. The localisations of the horizontal-flow inversions are better in general. Even at the depth of 17~Mm, the inversions can be said to be inversions at the target depth.

We detect systematic flows in the longitudinal direction. The systematic flows are related to the changes in rotation rate with depth. The comparison is plotted in Fig.~\ref{fig:vx_trend}. The correlation coefficient of the systematic flows and the changes in rotation rate are larger than 0.95 and the typical difference between the two is a few~\mps{}. This discrepancy is mostly caused by the fact that the detection algorithm averages the flows over a few degrees in latitude, while the model of rotation is purely equatorial. This independently confirms the validity of our inversions for the horizontal flows. 

We used the consistency of the horizontal flows and inverted them over a large range of depths down to 25~Mm. Therefore, the vertical flows computed from the continuity equation must be consistent too. The strong divergent horizontal flows become weaker with increasing depth and vanish at depths greater than about 7~Mm. We clearly identify the starting upflows at a depth of about 15~Mm below the surface. The upflows grow up to 3~Mm with amplitude of about 35~\mps{}. At shallower depth, the amplitude of the upflows decreases to the surface value of about 5~\mps{}. From this analysis, it seems that the average supergranule is clearly distinguishable from the background down to 10 Mm.


\begin{acknowledgements}
D.K. is supported by the Grant Agency of Charles University under grant No.~532217. M.Š. is supported by the project RVO:67985815. M.Š. and D.K. are supported by the grant project 18-06319S awarded by the Czech Science Foundation. The sensitivity kernels were computed by the code {\sc Kc3} kindly provided by Aaron Birch. This research has made use of NASA Astrophysics Data System Bibliographic Services. The authors would like to thank the anonymous referee for valuable comments and suggestions, which greatly improved the quality of the paper.
\end{acknowledgements}

\bibliographystyle{aa}
\bibliography{average_supergranule}


\begin{appendix}
\onecolumn

\section{The foreshortening-subtraction procedure}
\label{app:trend}

The mean travel times are spoiled with a centre-to-limb systematic trend. This trend is mostly caused by foreshortening due to the plane-parallel approximation. In this work, the Postel's projection is utilised. The travel times were measured around the disk centre, and therefore the foreshortening is a function of the distance from the disc centre. In the following list, we describe the procedure of the trend reduction.
\begin{enumerate}
    \item We compute the long-term (35~days) average of the measured mean travel times.
    \item We blur the averaged mean travel times with a wide Gaussian ($\FWHM{} \approx 30$~Mm) to remove all small-scale structures.
    \item Then we force the rotational symmetry of the averaged blurred mean travel times because it is reasonable to assume that the foreshortening is a function of the distance from the disc centre only.
    \item We then compute medians of concentric annuli whose centres are in the centre of the field of view and replace the elements of the averaged blurred mean travel times with the medians.
    \item We further blur the resulted images with the narrow Gaussian ($\FWHM{} \approx 3$~Mm) to remove discontinuities. 
    \item We subtract the values at the centre from the final images because we may safely assume that the foreshortening is exactly zero at the disc centre.
\end{enumerate}
In the left part of Fig.~\ref{fig:tt_mn_trend}, one can see a mean travel time after the trend is removed (before the supergranule alignment) and in the right part the corresponding trend is seen.

\tcfigure[!ht]{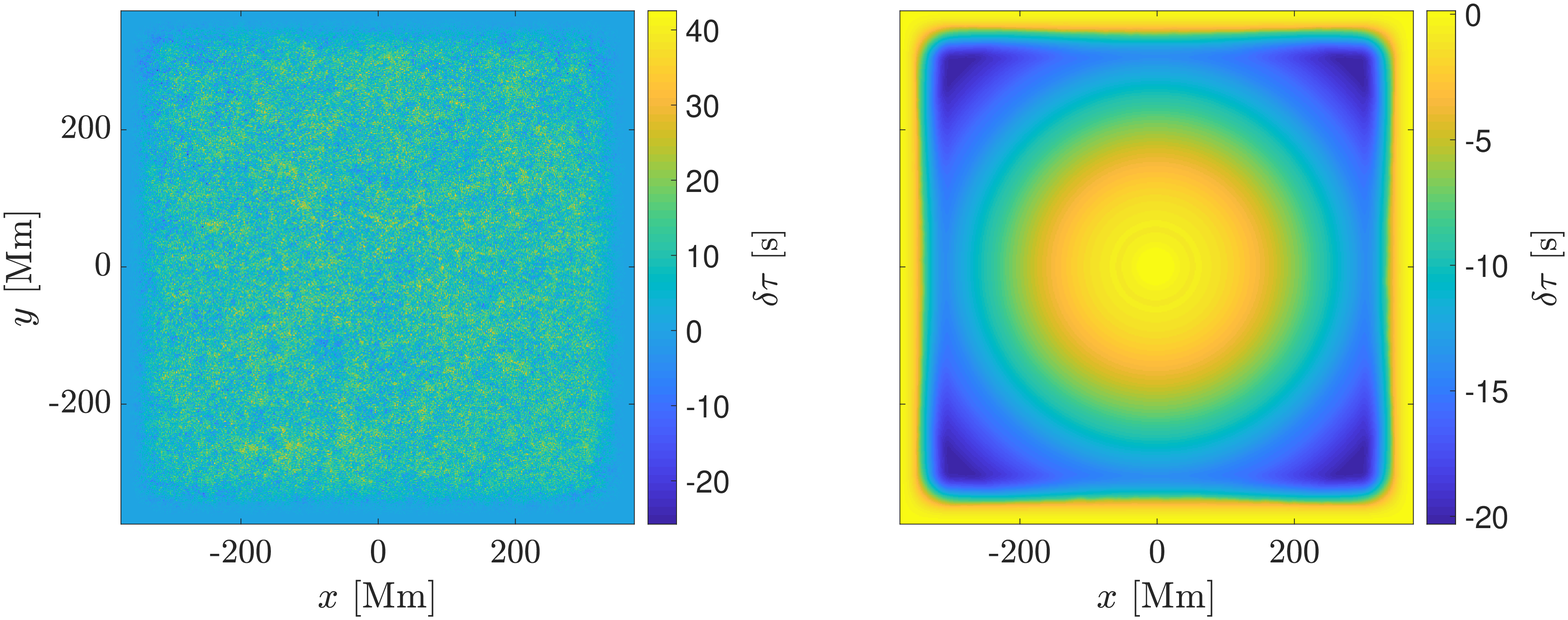}
{Mean $f$-mode travel time with \mbox{$\Delta \approx 16$~Mm}. Left: Travel time after the trend is subtracted. Right: Corresponding trend. }
{fig:tt_mn_trend}

\end{appendix}

\end{document}